\documentstyle[10pt,epsf,epsfig,hangcaption,xspace,amssymb,amsfonts,amsmath,amsthm,cite,
dp_delphititle,mathrsfs]{dp_delphi}
\makeindex
\def\DpPaperGroup{PH-EP}
\def\DpPaperRef{2004-067}
\def\DpDate{21 October 2004}
\def\DpAuthors{DELPHI Collaboration}
\def\DpTitle{{
Search for excited leptons in  {$\boldmath{\boldsymbol e^+ e^-}$} \\ 
collisions at {$\boldmath{\boldsymbol{\sqrt{s}=189-209}}$}~GeV
}}
\def\DpSubmit{(Accepted by Euro. Phys. Jour. C)}
\def\DpComment{}
\def\DpEMail{  }

\include{newcom}
\begin{document}
%%%%%%%%%%%%%%%%%%%%%%%%%% They are a problem with Coll.Sty ?
\makeatletter
%\input{dp_system:coll.sty}
% Collapse citation numbers to ranges.  Non-numeric and undefined labels
% are handled.  No sorting is done.  E.g., 1,3,2,3,4,5,foo,1,2,3,?,4,5
% gives 1,3,2-5,foo,1-3,?,4,5
\newcount\@tempcntc
\def\@citex[#1]#2{\if@filesw\immediate\write\@auxout{\string\citation{#2}}\fi
  \@tempcnta\z@\@tempcntb\m@ne\def\@citea{}\@cite{\@for\@citeb:=#2\do
    {\@ifundefined
       {b@\@citeb}{\@citeo\@tempcntb\m@ne\@citea\def\@citea{,}{\bf ?}\@warning
       {Citation `\@citeb' on page \thepage \space undefined}}%
    {\setbox\z@\hbox{\global\@tempcntc0\csname b@\@citeb\endcsname\relax}%
     \ifnum\@tempcntc=\z@ \@citeo\@tempcntb\m@ne
       \@citea\def\@citea{,}\hbox{\csname b@\@citeb\endcsname}%
     \else
      \advance\@tempcntb\@ne
      \ifnum\@tempcntb=\@tempcntc
      \else\advance\@tempcntb\m@ne\@citeo
      \@tempcnta\@tempcntc\@tempcntb\@tempcntc\fi\fi}}\@citeo}{#1}}
\def\@citeo{\ifnum\@tempcnta>\@tempcntb\else\@citea\def\@citea{,}%
  \ifnum\@tempcnta=\@tempcntb\the\@tempcnta\else
   {\advance\@tempcnta\@ne\ifnum\@tempcnta=\@tempcntb \else \def\@citea{--}\fi
    \advance\@tempcnta\m@ne\the\@tempcnta\@citea\the\@tempcntb}\fi\fi}
 
\makeatother
%%%%%%%%%%%%%%%%%%%%%%%%%% ??????????????????????????????????
% Generate the title page
\begin{titlepage}
\pagenumbering{roman}
\CERNpreprint{\DpPaperGroup}{\DpPaperRef} % Reference of the paper
\date{{\small\DpDate}} % Date of the paper
\title{\DpTitle} % Title of the paper
\address{\DpAuthors} % General name of the author(s)
\begin{shortabs} % Start the abstract
\noindent
%===================> DELPHI note abstract     =====> To be filled <=====%
A search for excited lepton production in $e^+ e^-$ collisions was 
performed using the data collected by the DELPHI detector at LEP  
at centre-of-mass energies ranging from 189~GeV to 209~GeV,
corresponding to an integrated luminosity of 
approximately 600~pb$^{-1}$. No evidence for excited lepton production  
was found.
In searches for pair-produced excited leptons, lower mass limits were
established in the range \mbox{94 -- 103~GeV/$c^2$}, depending on the channel
and model assumptions. 
In searches for singly-produced excited leptons, upper limits on the 
parameter $f/\Lambda$ were established as a function of the mass.
%% (D.C.) All limits are at 95\% confidence level.
\end{shortabs}
\vfill
\begin{center}
\DpSubmit \ \\ % Horrible hack to allow to have DpSubmit empty
\DpComment \ \\
\DpEMail \ \\
\end{center}
\vfill
\clearpage
\headsep 10.0pt
\addtolength{\textheight}{10mm}
\addtolength{\footskip}{-5mm}
\begingroup
% Commands to process the author names
%
\newcommand{\DpName}[2]{\hbox{#1$^{\ref{#2}}$},\hfill}
\newcommand{\DpNameTwo}[3]{\hbox{#1$^{\ref{#2},\ref{#3}}$},\hfill}
\newcommand{\DpNameThree}[4]{\hbox{#1$^{\ref{#2},\ref{#3},\ref{#4}}$},\hfill}
\newskip\Bigfill \Bigfill = 0pt plus 1000fill
\newcommand{\DpNameLast}[2]{\hbox{#1$^{\ref{#2}}$}\hspace{\Bigfill}}
%
%\small
\footnotesize
\noindent
\DpName{J.Abdallah}{LPNHE}
\DpName{P.Abreu}{LIP}
\DpName{W.Adam}{VIENNA}
\DpName{P.Adzic}{DEMOKRITOS}
\DpName{T.Albrecht}{KARLSRUHE}
\DpName{T.Alderweireld}{AIM}
\DpName{R.Alemany-Fernandez}{CERN}
\DpName{T.Allmendinger}{KARLSRUHE}
\DpName{P.P.Allport}{LIVERPOOL}
\DpName{U.Amaldi}{MILANO2}
% ...
\normalsize
\endgroup
\titlefoot{Department of Physics and Astronomy, Iowa State
     University, Ames IA 50011-3160, USA
    \label{AMES}}
\titlefoot{Physics Department, Universiteit Antwerpen,
     Universiteitsplein 1, B-2610 Antwerpen, Belgium \\
     \indent~~and IIHE, ULB-VUB,
     Pleinlaan 2, B-1050 Brussels, Belgium \\
     \indent~~and Facult\'e des Sciences,
     Univ. de l'Etat Mons, Av. Maistriau 19, B-7000 Mons, Belgium
    \label{AIM}}
\titlefoot{Physics Laboratory, University of Athens, Solonos Str.
     104, GR-10680 Athens, Greece
    \label{ATHENS}}
\titlefoot{Department of Physics, University of Bergen,
     All\'egaten 55, NO-5007 Bergen, Norway
    \label{BERGEN}}
\titlefoot{Dipartimento di Fisica, Universit\`a di Bologna and INFN,
     Via Irnerio 46, IT-40126 Bologna, Italy
    \label{BOLOGNA}}
\titlefoot{Centro Brasileiro de Pesquisas F\'{\i}sicas, rua Xavier Sigaud 150,
     BR-22290 Rio de Janeiro, Brazil \\
     \indent~~and Depto. de F\'{\i}sica, Pont. Univ. Cat\'olica,
     C.P. 38071 BR-22453 Rio de Janeiro, Brazil \\
     \indent~~and Inst. de F\'{\i}sica, Univ. Estadual do Rio de Janeiro,
     rua S\~{a}o Francisco Xavier 524, Rio de Janeiro, Brazil
    \label{BRASIL}}
\titlefoot{Coll\`ege de France, Lab. de Physique Corpusculaire, IN2P3-CNRS,
     FR-75231 Paris Cedex 05, France
    \label{CDF}}
\titlefoot{CERN, CH-1211 Geneva 23, Switzerland
    \label{CERN}}
\titlefoot{Institut de Recherches Subatomiques, IN2P3 - CNRS/ULP - BP20,
     FR-67037 Strasbourg Cedex, France
    \label{CRN}}
\titlefoot{Now at DESY-Zeuthen, Platanenallee 6, D-15735 Zeuthen, Germany
    \label{DESY}}
\titlefoot{Institute of Nuclear Physics, N.C.S.R. Demokritos,
     P.O. Box 60228, GR-15310 Athens, Greece
    \label{DEMOKRITOS}}
\titlefoot{FZU, Inst. of Phys. of the C.A.S. High Energy Physics Division,
     Na Slovance 2, CZ-180 40, Praha 8, Czech Republic
    \label{FZU}}
\titlefoot{Dipartimento di Fisica, Universit\`a di Genova and INFN,
     Via Dodecaneso 33, IT-16146 Genova, Italy
    \label{GENOVA}}
\titlefoot{Institut des Sciences Nucl\'eaires, IN2P3-CNRS, Universit\'e
     de Grenoble 1, FR-38026 Grenoble Cedex, France
    \label{GRENOBLE}}
\titlefoot{Helsinki Institute of Physics, P.O. Box 64,
     FIN-00014 University of Helsinki, Finland
    \label{HELSINKI}}
\titlefoot{Joint Institute for Nuclear Research, Dubna, Head Post
     Office, P.O. Box 79, RU-101 000 Moscow, Russian Federation
    \label{JINR}}
\titlefoot{Institut f\"ur Experimentelle Kernphysik,
     Universit\"at Karlsruhe, Postfach 6980, DE-76128 Karlsruhe,
     Germany
    \label{KARLSRUHE}}
\titlefoot{Institute of Nuclear Physics PAN,Ul. Radzikowskiego 152,
     PL-31142 Krakow, Poland
    \label{KRAKOW1}}
\titlefoot{Faculty of Physics and Nuclear Techniques, University of Mining
     and Metallurgy, PL-30055 Krakow, Poland
    \label{KRAKOW2}}
\titlefoot{Universit\'e de Paris-Sud, Lab. de l'Acc\'el\'erateur
     Lin\'eaire, IN2P3-CNRS, B\^{a}t. 200, FR-91405 Orsay Cedex, France
    \label{LAL}}
\titlefoot{School of Physics and Chemistry, University of Lancaster,
     Lancaster LA1 4YB, UK
    \label{LANCASTER}}
\titlefoot{LIP, IST, FCUL - Av. Elias Garcia, 14-$1^{o}$,
     PT-1000 Lisboa Codex, Portugal
    \label{LIP}}
\titlefoot{Department of Physics, University of Liverpool, P.O.
     Box 147, Liverpool L69 3BX, UK
    \label{LIVERPOOL}}
\titlefoot{Dept. of Physics and Astronomy, Kelvin Building,
     University of Glasgow, Glasgow G12 8QQ, UK.
    \label{GLASGOW}}
\titlefoot{LPNHE, IN2P3-CNRS, Univ.~Paris VI et VII, Tour 33 (RdC),
     4 place Jussieu, FR-75252 Paris Cedex 05, France
    \label{LPNHE}}
\titlefoot{Department of Physics, University of Lund,
     S\"olvegatan 14, SE-223 63 Lund, Sweden
    \label{LUND}}
\titlefoot{Universit\'e Claude Bernard de Lyon, IPNL, IN2P3-CNRS,
     FR-69622 Villeurbanne Cedex, France
    \label{LYON}}
\titlefoot{Dipartimento di Fisica, Universit\`a di Milano and INFN-MILANO,
     Via Celoria 16, IT-20133 Milan, Italy
    \label{MILANO}}
\titlefoot{Dipartimento di Fisica, Univ. di Milano-Bicocca and
     INFN-MILANO, Piazza della Scienza 2, IT-20126 Milan, Italy
    \label{MILANO2}}
\titlefoot{IPNP of MFF, Charles Univ., Areal MFF,
     V Holesovickach 2, CZ-180 00, Praha 8, Czech Republic
    \label{NC}}
\titlefoot{NIKHEF, Postbus 41882, NL-1009 DB
     Amsterdam, The Netherlands
    \label{NIKHEF}}
\titlefoot{National Technical University, Physics Department,
     Zografou Campus, GR-15773 Athens, Greece
    \label{NTU-ATHENS}}
\titlefoot{Physics Department, University of Oslo, Blindern,
     NO-0316 Oslo, Norway
    \label{OSLO}}
\titlefoot{Dpto. Fisica, Univ. Oviedo, Avda. Calvo Sotelo
     s/n, ES-33007 Oviedo, Spain
    \label{OVIEDO}}
\titlefoot{Department of Physics, University of Oxford,
     Keble Road, Oxford OX1 3RH, UK
    \label{OXFORD}}
\titlefoot{Dipartimento di Fisica, Universit\`a di Padova and
     INFN, Via Marzolo 8, IT-35131 Padua, Italy
    \label{PADOVA}}
\titlefoot{Rutherford Appleton Laboratory, Chilton, Didcot
     OX11 OQX, UK
    \label{RAL}}
\titlefoot{Dipartimento di Fisica, Universit\`a di Roma II and
     INFN, Tor Vergata, IT-00173 Rome, Italy
    \label{ROMA2}}
\titlefoot{Dipartimento di Fisica, Universit\`a di Roma III and
     INFN, Via della Vasca Navale 84, IT-00146 Rome, Italy
    \label{ROMA3}}
\titlefoot{DAPNIA/Service de Physique des Particules,
     CEA-Saclay, FR-91191 Gif-sur-Yvette Cedex, France
    \label{SACLAY}}
\titlefoot{Instituto de Fisica de Cantabria (CSIC-UC), Avda.
     los Castros s/n, ES-39006 Santander, Spain
    \label{SANTANDER}}
\titlefoot{Inst. for High Energy Physics, Serpukov
     P.O. Box 35, Protvino, (Moscow Region), Russian Federation
    \label{SERPUKHOV}}
\titlefoot{J. Stefan Institute, Jamova 39, SI-1000 Ljubljana, Slovenia
     and Laboratory for Astroparticle Physics,\\
     \indent~~Nova Gorica Polytechnic, Kostanjeviska 16a, SI-5000 Nova Gorica, Slovenia, \\
     \indent~~and Department of Physics, University of Ljubljana,
     SI-1000 Ljubljana, Slovenia
    \label{SLOVENIJA}}
\titlefoot{Fysikum, Stockholm University,
     Box 6730, SE-113 85 Stockholm, Sweden
    \label{STOCKHOLM}}
\titlefoot{Dipartimento di Fisica Sperimentale, Universit\`a di
     Torino and INFN, Via P. Giuria 1, IT-10125 Turin, Italy
    \label{TORINO}}
\titlefoot{INFN,Sezione di Torino, and Dipartimento di Fisica Teorica,
     Universit\`a di Torino, Via P. Giuria 1,\\
     \indent~~IT-10125 Turin, Italy
    \label{TORINOTH}}
\titlefoot{Dipartimento di Fisica, Universit\`a di Trieste and
     INFN, Via A. Valerio 2, IT-34127 Trieste, Italy \\
     \indent~~and Istituto di Fisica, Universit\`a di Udine,
     IT-33100 Udine, Italy
    \label{TU}}
\titlefoot{Univ. Federal do Rio de Janeiro, C.P. 68528
     Cidade Univ., Ilha do Fund\~ao
     BR-21945-970 Rio de Janeiro, Brazil
    \label{UFRJ}}
\titlefoot{Department of Radiation Sciences, University of
     Uppsala, P.O. Box 535, SE-751 21 Uppsala, Sweden
    \label{UPPSALA}}
\titlefoot{IFIC, Valencia-CSIC, and D.F.A.M.N., U. de Valencia,
     Avda. Dr. Moliner 50, ES-46100 Burjassot (Valencia), Spain
    \label{VALENCIA}}
\titlefoot{Institut f\"ur Hochenergiephysik, \"Osterr. Akad.
     d. Wissensch., Nikolsdorfergasse 18, AT-1050 Vienna, Austria
    \label{VIENNA}}
\titlefoot{Inst. Nuclear Studies and University of Warsaw, Ul.
     Hoza 69, PL-00681 Warsaw, Poland
    \label{WARSZAWA}}
\titlefoot{Fachbereich Physik, University of Wuppertal, Postfach
     100 127, DE-42097 Wuppertal, Germany
    \label{WUPPERTAL}}
\addtolength{\textheight}{-10mm}
\addtolength{\footskip}{5mm}
\clearpage
\headsep 30.0pt
\end{titlepage}
%%%%%%%%%%%%%%%%%%%%%%%%%
%
% Change for the document body
%%\pagestyle{heading} % for page numbering
\pagenumbering{arabic} % page numbering in number
\setcounter{footnote}{0} %
\large
%\linenumbers %%%CD
%%%%%%%%%%%%%%%%%%%%%%%%%%%%%%%%%%%%%%%%%%%%%%%%%%%%%%%%%%%%%%%%%%%%%%%%%%%%%%%
\def\asmz{$\alpha_s(M_Z)$}
\def\ass{\alpha_s(\sqrt{s})}
\newcommand{\kos}{\ifmmode {{\mathrm K}^{0}_{S}} \else
${\mathrm K}^{0}_{S}$\fi}
\newcommand{\kpm}{\ifmmode {{\mathrm K}^{\pm}} \else
${\mathrm K}^{\pm}$\fi}
\newcommand{\ko}{\ifmmode {{\mathrm K}^{0}} \else
${\mathrm K}^{0}$\fi}

\newcommand{\wboson}{\ifmmode {{\mathrm W}^{\pm}} \else
${\mathrm W}^{\pm}$\fi}
\newcommand{\wpair}{\ifmmode {{\mathrm W}^{+}{\mathrm W}^{-} } \else
${\mathrm W}^{+}{\mathrm W}^{-}$\fi}
\newcommand{\zboson}{\ifmmode {{\mathrm Z}^{0}} \else
${\mathrm Z}^{0}$\fi}

\newcommand{\elepton}{\ifmmode {L^{*}} \else $L^{*}$\fi}
\newcommand{\lepton}{\ifmmode {L}      \else $L$\fi}

\newcommand{\cme}{centre-of-mass energy}
\newcommand{\cmes}{centre-of-mass energies}

\def\as{$\alpha_s$}
\def\asb{$\alpha_s\sp{b}$}

\def\asc{$\alpha_s\sp{c}$}
\def\asuds{$\alpha_s\sp{udsc}$}
\def\Lam{$\Lambda$}
\def\ZP{Zeit.\ Phys.\ {\bf C}}
\def\PL{Phys.\ Lett.\ {\bf B}}
\def\PR{Phys.\ Rev.\ {\bf D}}
\def\PRL{Phys.\ Rev.\ Lett.\ }
\def\NP{Nucl.\ Phys.\ {\bf B}}
\def\CPC{Comp.\ Phys.\ Comm.\ }
\def\NIM{Nucl.\ Instr.\ Meth.\ }
\def\Coll{Coll.,\ }
\def\Rmu{$R_3(\mu)/R_3(had)$\ }
\def\Re{$R_3(e)/R_3(had)$\ }
\def\Rmue{$R_3(\mu +e)/R_3(had)$\ }
\def\ee{$e\sp{+}e\sp{-}$}

\def\pslash{\not{\hbox{\kern-2.3pt$p$}}}

\def\eslash{\not{\hbox{\kern-3.5pt$E$}}}

\section{Introduction}
\label{sec:intro}

Excited leptons are predicted by models with substructure in the 
fermionic sector~\cite{CERNreports,hagiwara,boudjema}.   
They  are assumed to have spin and weak isospin equal 
to 1/2 and to have both their left-handed and right-handed components 
arranged in weak isodoublets:

%% $$
%% {L}^*_L = \begin{bmatrix} \nu^* \\ \ell^*  \end{bmatrix}_L ; \quad 
%% {L}^*_R = \begin{bmatrix} \nu^* \\ \ell^*  \end{bmatrix}_R  
%% $$

$$
{L}^*_L = \binom{\nu^*}{\ell^*}_L \quad  {L}^*_R = \binom{\nu^*}{\ell^*}_R  
$$

\noindent
where $\nu^* = \nu^*_e,\nu^*_\mu,\nu^*_\tau$ and  $\ell^* = e^*, \mu^*, \tau^*$ represent 
the different flavours of neutral and charged excited leptons respectively.
Excited leptons (\mbox{$\elepton \equiv \ell^*, \nu^*$}) couple to the photon and/or to the 
\wboson\ and \zboson\  gauge bosons according to their quantum numbers and thus could be 
pair produced at LEP. 
Single production in association with their Standard Model (SM) partners 
($\lepton \equiv \ell, \nu$) would also be possible and its cross-sections 
would depend on the $\lepton \elepton V$ couplings 
\mbox{($V \equiv \gamma,\wboson,\zboson$)} \cite{djouadi}. 
Excited leptons with masses up to the \cme\ ($\sqrt{s}$) can be searched for through the 
single-production mode.

This paper presents results of a search for single and pair production 
of excited leptons of all flavours using data collected by the DELPHI experiment 
at LEP at centre-of-mass energies between 189~GeV and 209~GeV.
Previous results by DELPHI and the other LEP experiments can be
found in references~\cite{delphi1,delphi2,LEP}, while results from
the HERA experiments can be found in reference~\cite{HERA}. 
The paper is organised as follows: 
section~\ref{sec:theo} reviews the phenomenology of excited-lepton production and decay 
and its consequences for the experimental strategy;  
in section~\ref{sec:detect} a detector overview is given and the data and Monte Carlo 
simulations are presented;
event selection criteria applied  to the different search channels 
are described in section~\ref{sec:selection} and the results obtained are 
presented in section~\ref{sec:results}; 
finally, a  summary is presented in section~\ref{sec:summary}. 

\section{Production and decay of excited leptons}
\label{sec:theo}

\begin{figure}[p]
\begin{center}
\vspace{-0.5cm}
\mbox{\epsfig{file=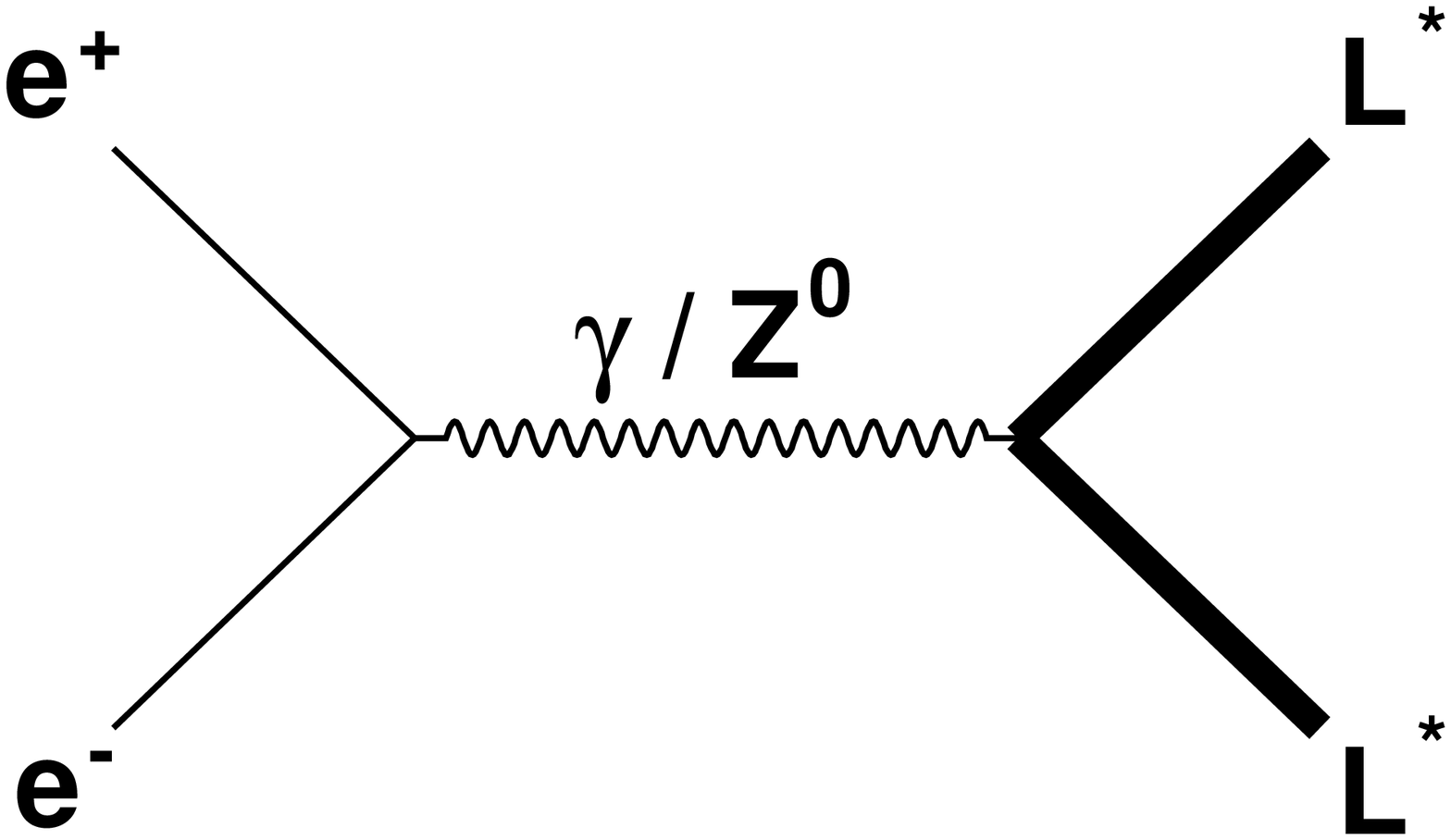,width=0.3\textwidth}}      \hspace{0.03\textwidth}
\mbox{\epsfig{file=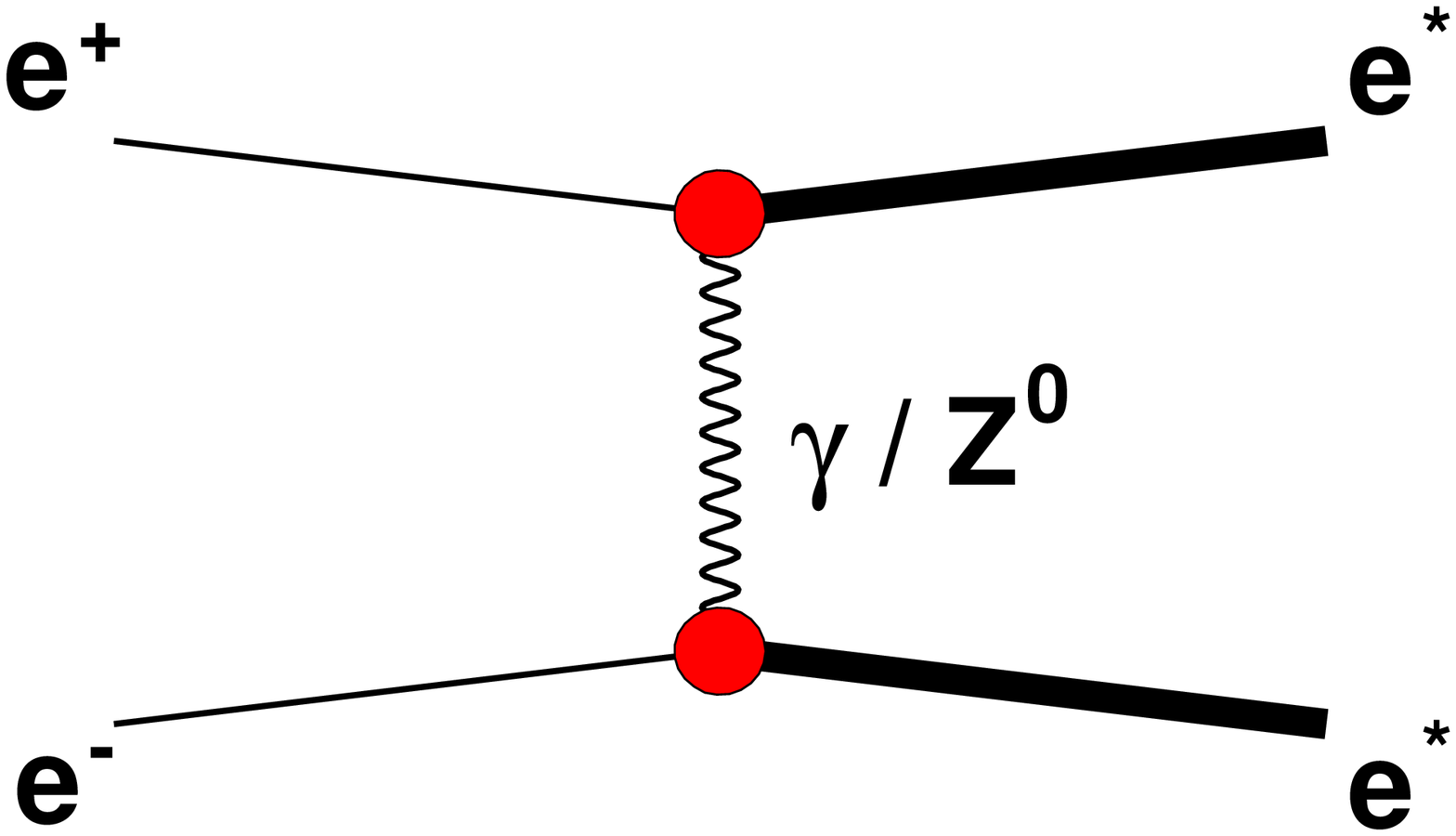,width=0.3\textwidth}} \hspace{0.03\textwidth}
\mbox{\epsfig{file=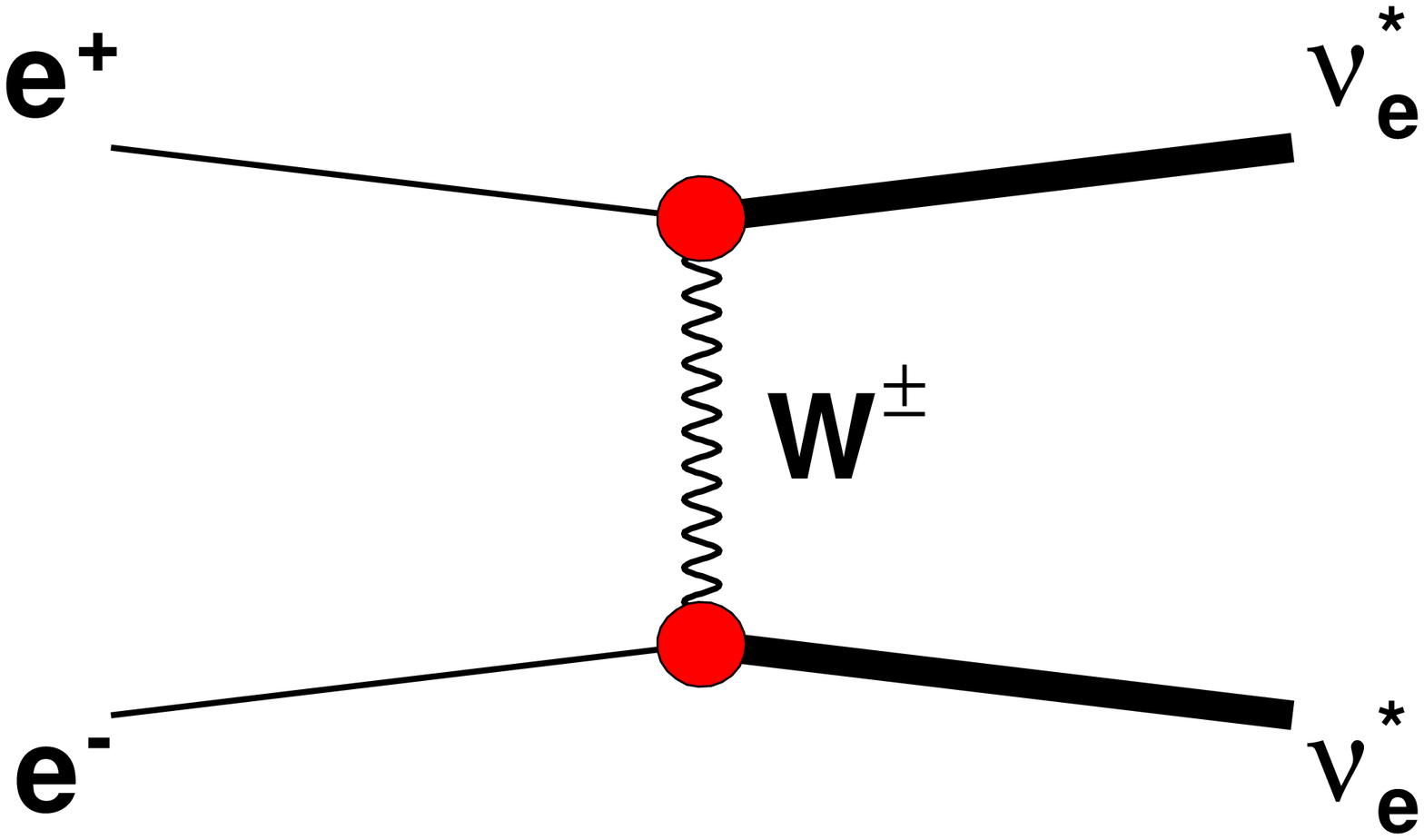,width=0.3\textwidth}}

\vspace{0.5cm}

\mbox{\epsfig{file=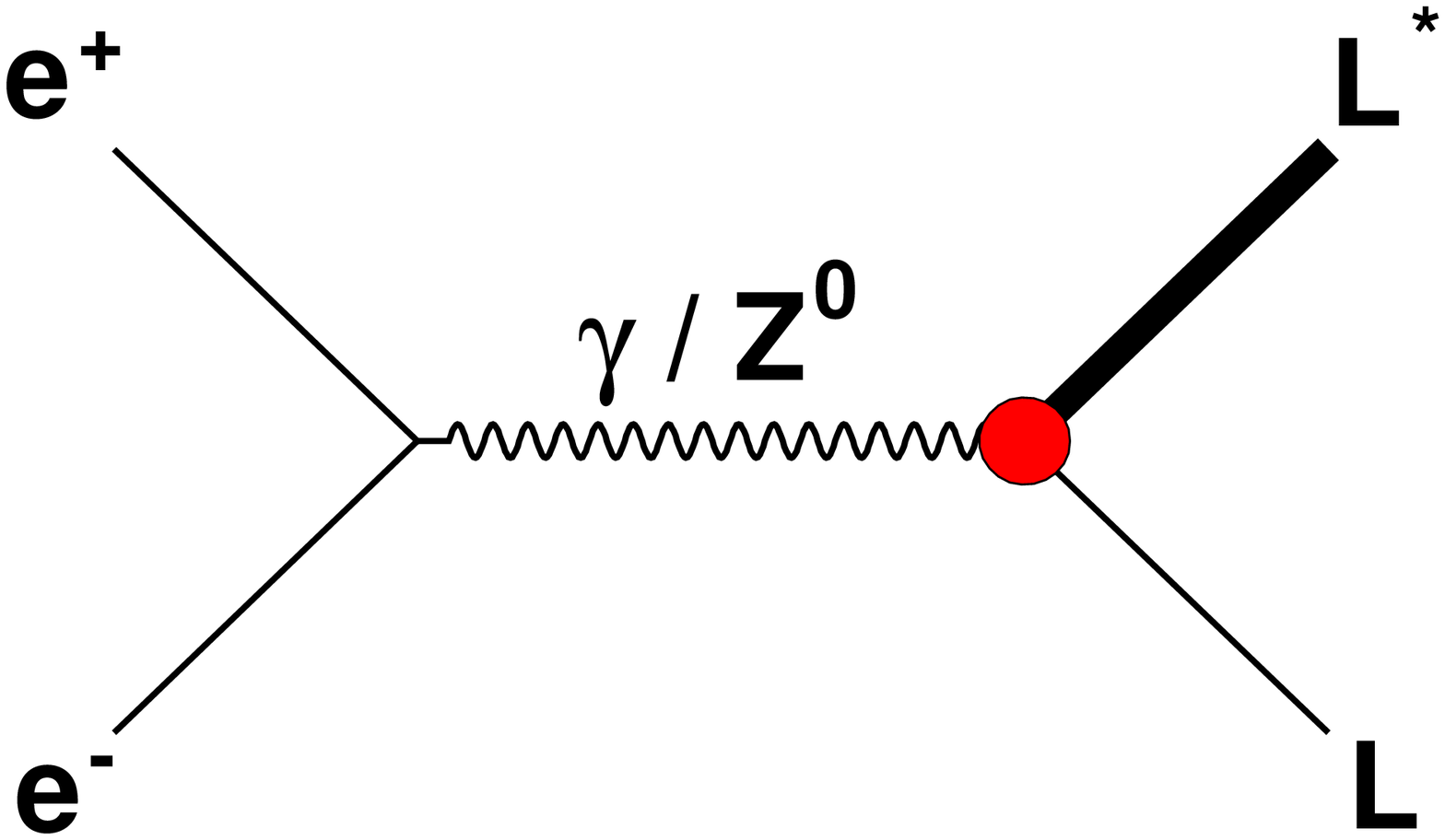,width=0.3\textwidth}}      \hspace{0.03\textwidth}
\mbox{\epsfig{file=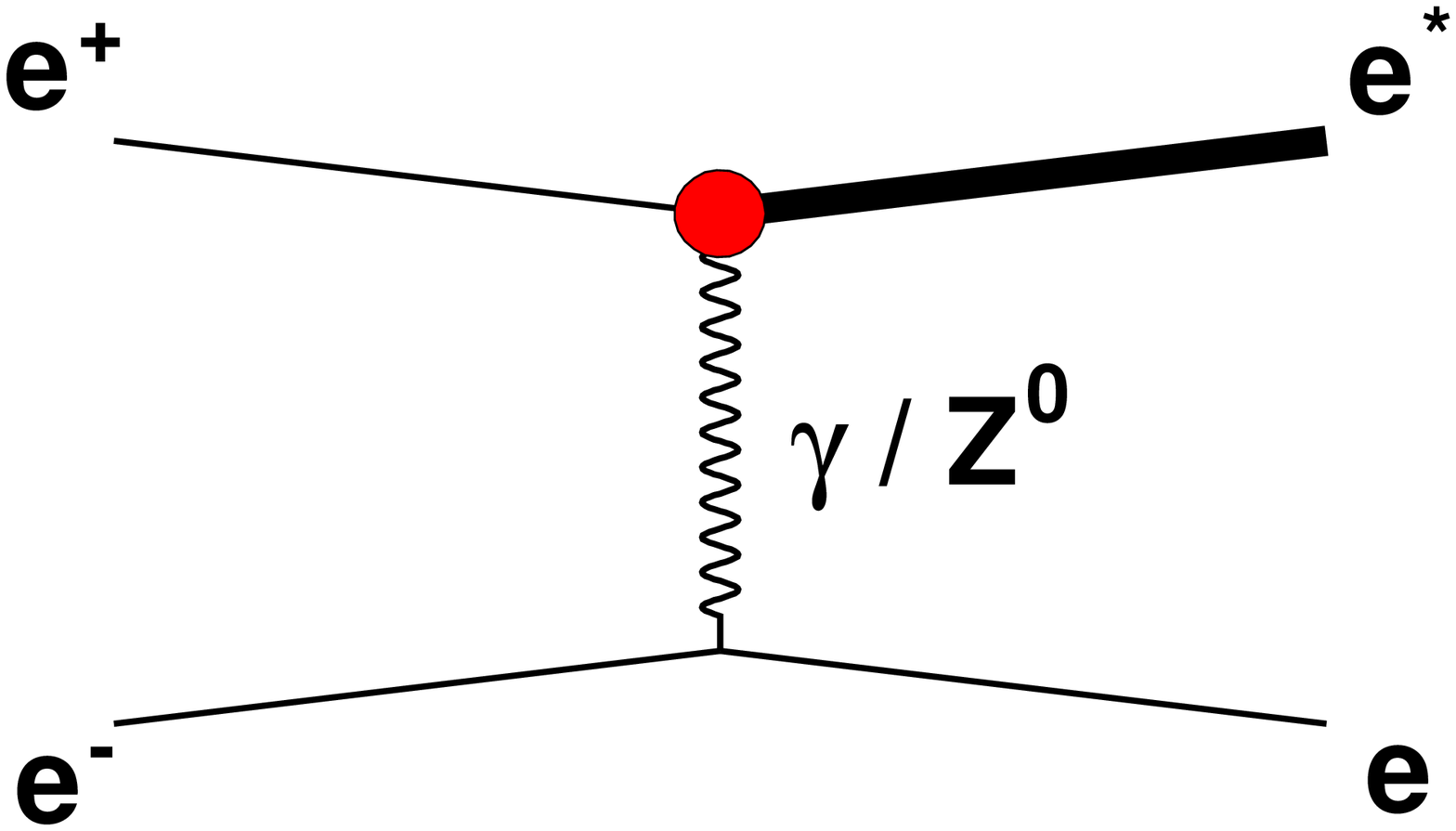,width=0.3\textwidth}} \hspace{0.03\textwidth}
\mbox{\epsfig{file=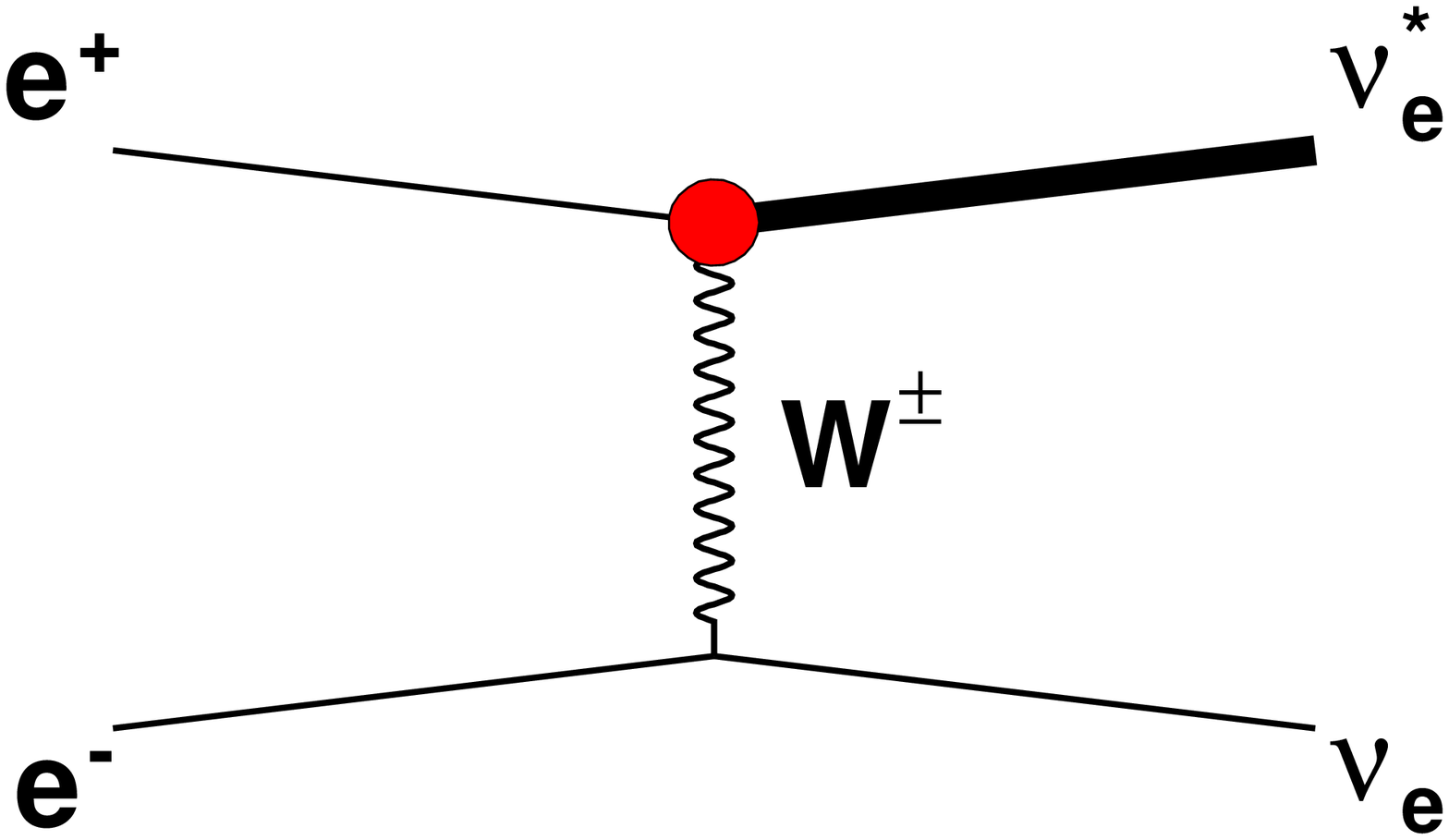,width=0.3\textwidth}}

\caption{Feynman diagrams for the double (top) and single (bottom) excited-lepton 
production. Each \elepton\ is shown as a thicker line.
The vertex shown as a closed circle represents a $L \elepton V$ coupling
(\mbox{$V \equiv \gamma,\wboson,\zboson$}) inversely proportional to the 
compositeness scale parameter $\Lambda$.}
\label{fig:Feynman}
\end{center}
\end{figure}

Pair production of charged excited leptons in $e^+ e^-$ collisions  
proceeds via $s$-channel $\gamma$ and \zboson\ exchanges, 
while for excited neutrinos only the \zboson\ diagram contributes 
(figure~\ref{fig:Feynman}). 
Pair production of excited electrons or excited electron neutrinos is
also possible through \mbox{$t$-channel} exchange diagrams. 
In this case two $\lepton \elepton V$ vertices are involved and the contribution 
to the total production cross-section is expected to be negligible compared 
to the $s$-channel exchange diagrams. 

Single excited-lepton production arises from  $s$-channel photon and \zboson\ 
exchange (figure~\ref{fig:Feynman}). 
Important additional contributions from $t$-channel $\gamma$ and
\zboson\ exchange arise for excited electron production, while 
$t$-channel \wboson\ exchange can be important for the excited electron neutrino.
In the $t$-channel production of excited electrons, the SM spectator electron is 
emitted at small angles to the colliding beams direction and thus is often not detected. 
%%% goes undetected in the beam pipe. 
 
The SU(2)$\times$U(1) gauge-invariant effective  Lagrangian describing the  
magnetic transition between excited leptons and the SM leptons has the 
form~\cite{boudjema}: 

$$
 {\mathscr L}_{\lepton \elepton} = 
                   \frac{1}{2\Lambda} \overline{{L}^*} 
                    \sigma^{\mu \nu} 
                   \left[ 
                         g f \frac{{\overrightarrow \tau}}{2} \cdot {\overrightarrow W}_{\mu \nu} +
                         g' f' \frac{Y}{2} B_{\mu \nu} 
                   \right] {L}_L       
                   + \text{hermitian conjugate}
$$

\noindent
where ${L}^*={L}^*_L + {L}^*_R $ and $L_L$ is the weak isodoublet with the 
left-handed components of the SM leptons;
$\sigma^{\mu \nu}$ is the covariant bilinear tensor, 
$\overrightarrow  \tau$ are the Pauli matrices,  $Y$ is the weak hypercharge,  
${\overrightarrow  W}_{\mu \nu}$ and $B_{\mu \nu}$ represent the gauge field tensors of 
SU(2) and U(1) respectively, with $g$ and $g'$ being the corresponding 
SM coupling constants; 
the parameter $\Lambda$ sets the compositeness scale, with 
$f$ and $f'$ being weight factors associated with the two gauge groups.
%% A more extensive discussion of this effective Lagrangian can be found 
%% in reference~\cite{hagiboud}. 
This Lagrangian is associated to the $\lepton \elepton V$ vertex, and describes
the single production of excited leptons and their decay branching ratios radiating
a photon or a \wboson\ or \zboson\ boson.
The strength of the $\lepton \elepton V$ coupling is parameterised through 
$f$ and $f'$.
Form factors and anomalous magnetic moments of the excited leptons were not
considered in the reported analyses.
To reduce the number of free parameters it is customary to consider a relation between $f$
and $f'$, or set one of the parameters to zero. In this paper the relations $f=f'$ and
$f=-f'$ are assumed. 
With the assumption $|f|=|f'|$, 
%%%% or assuming that only one of the constants $f$ is non-negligible,        %%%
the single excited-lepton production cross-section depends only on the ratio 
$f/\Lambda$ and on the excited-lepton mass.
%The search results can also be expressed in terms of the parameter $\lambda$,
%which is related  to the excited lepton mass ($m_{\elepton}$) through  
%$f/\Lambda = \sqrt{2} \lambda/m_{\elepton}$.

\begin{figure}[htb]
\begin{center}
\vspace{-1.cm}
\mbox{\epsfig{file=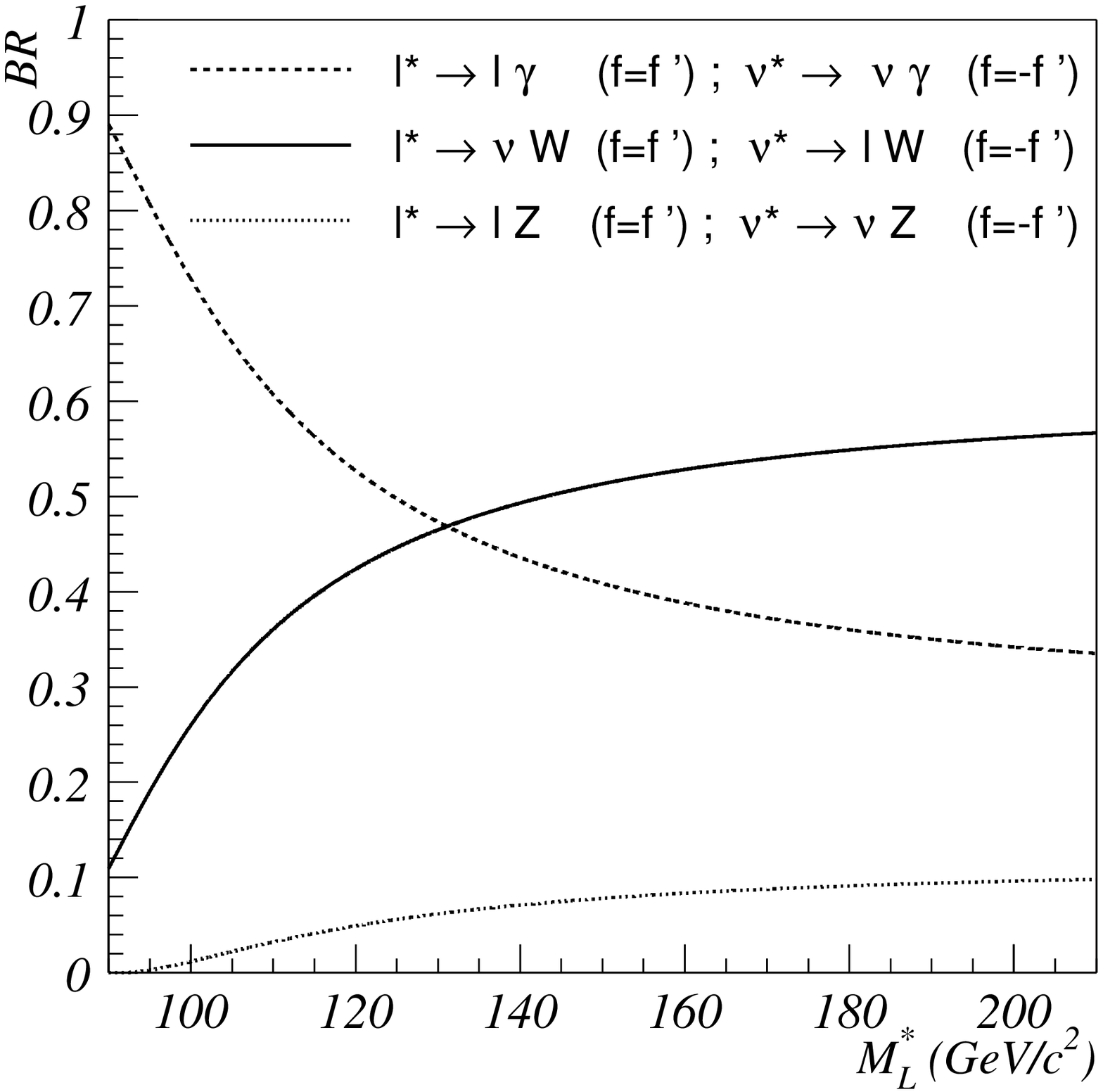,width=0.485\textwidth}}
\mbox{\epsfig{file=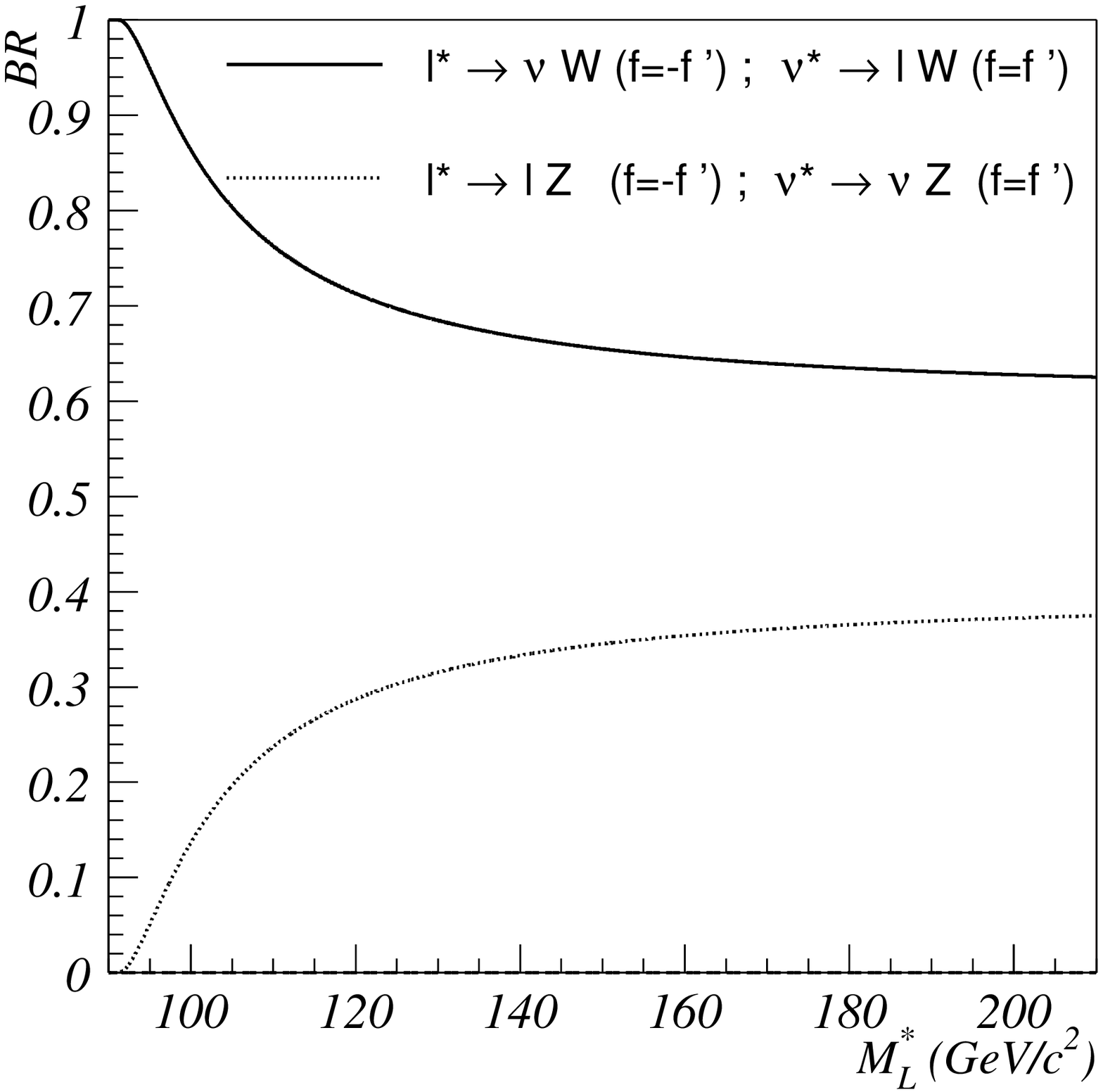,width=0.485\textwidth}}
%\vspace{0.5cm}
\caption{Branching ratios of the excited-lepton decays as a function of the mass.
In the left plot are shown the branching ratios for charged and neutral excited leptons when 
$f=f'$ and $f=-f'$, respectively.  
The right plot refers to the symmetric $f$ and $f'$ couplings assignment.  }
\label{fig:br}
\end{center}
\end{figure}

Excited leptons with masses above 20~GeV/$c^2$ are assumed to decay promptly 
by radiating a photon, \wboson\   or \zboson\ boson. 
Their mean lifetime is predicted to be less than $10^{-15}$~s in all the studied 
scenarios. Thus, for detection purposes, excited leptons essentially decay 
at the production point.
%%% thus decay lengths are about 1~cm or shorter.
%%%%
The decay branching ratios are function of the $f$ and $f'$ parameters, as
shown in figure~\ref{fig:br} as a function of the excited-lepton mass ($m_{\elepton}$)
for $f=f'$ and $f=-f'$.
For charged excited leptons, the electromagnetic radiative decay is forbidden
if \mbox{$f=-f'$} and the decays proceed exclusively 
through  \zboson\ and \wboson\ bosons.  
However, as long as \mbox{$f \neq -f'$} there is a significant contribution to the 
total decay width from the electromagnetic radiative decay, even if   
$|f|-|f'| \ll |f|$. 
If \mbox{$f=f'$}, the electromagnetic radiative decay 
branching ratio is close to $100$\% for $m_{\elepton}$ smaller than the \wboson\ mass
($m_{\mathrm W}$),  but decreases for $m_{\elepton} > m_{\mathrm W}$ reaching a 
value of 34\% for \mbox{$m_{\elepton}=200$~GeV/$c^2$}.
For excited neutrinos the electromagnetic decays are forbidden only if $f=f'$.

Many final-state topologies arise from the production and decay of excited leptons. 
They involve isolated leptons, isolated photons, particle jets from quark fragmentation, 
missing energy ($\eslash$) and missing momentum ($\pslash$). 

\begin{table}[hbt]
\begin{center}
%\vskip 0.45 cm
{
\begin{tabular}{|l|c|c|}
\hline
      & \multicolumn{2}{c|}{Final-state Topologies} 
\\ \cline{2-3}
Channel & Single production & Pair production
\\
\hline
\hline
$\ell^* \rightarrow \ell\gamma$ &  $\ell\ell\gamma$, $(\ell)\ell\gamma$  
& $\ell\ell\gamma\gamma$ \\ 
$\ell^* \rightarrow \nu W$      &  $jj\ell\!\eslash$, $jj(\ell)\!\eslash$ 
& $jj\ell\!\eslash$, $jjjj\!\eslash$ 
\\ 
$\ell^* \rightarrow \ell Z$     & $jj\ell\ell$, $jj\ell(\ell)$ & -
\\ \hline
$\nu^* \rightarrow \nu \gamma $ &  $\gamma\!\eslash$             
& $\gamma\gamma\!\eslash$
\\ 
$\nu^* \rightarrow \ell W$    &   $jj\ell\!\eslash$
&   $jj\ell\ell\ell$, $jj\ell\ell(\ell)$, $jjjj\ell\ell$  
\\ 
$\nu^* \rightarrow \nu Z$     &  $jj\!\eslash$         & -
\\ 
\hline           
\end{tabular}
}
\end{center}
\caption[]{Analysed final-state topologies corresponding to the different
production and decay modes of excited leptons. 
The spectator or final state SM lepton remaining undetected is indicated by ($\ell$).} 
\label{tab:topo}
\end{table}

\noindent
Table \ref{tab:topo} shows the topologies considered
for the different \elepton\ production and decay channels.
Several of those,  although not corresponding directly to the physical 
final state, are expected to become particularly important in presence of
low energy or low polar angle\footnote{In DELPHI a 
right-handed Cartesian coordinate system was used, with the $z$-axis pointing along 
the electron beam, the $x$-axis pointing toward the centre of the LEP ring and the origin 
at the centre of the detector. The polar angle $\theta$ is the angle 
to the electron beam direction and the azimuthal angle $\phi$ is the angle 
measured from the $x$-axis.
In this paper $\theta$ also refers to the complementary angle $180^\circ-\theta$.
} leptons. %%% For these topologies the unseen lepton is shown in parentheses.

The final states arising from the \wboson\ or \zboson\ leptonic decays 
in the single-production mode and 
the  \wpair\ purely leptonic decays or 
the \mbox{\zboson \zboson } final states in the pair-production searches 
were not considered due to their small branching ratio and/or 
small signal sensitivity. 
In addition the pair-production search criteria aimed at selecting only the final state 
topologies where both excited leptons decay to identical gauge bosons. 
(unmixed decays). Mixed decays where each excited lepton decays to a different gauge boson
are not considered.

\section{Detector overview and data samples}
\label{sec:detect}

The data analysed were collected with the DELPHI detector in the years \mbox{1998--2000}, 
at \cmes\  ranging from  189~GeV to 209~GeV and correspond to a total integrated 
luminosity of 598.7~pb$^{-1}$, with an average \cme\  of 
\mbox{$<\!\!\sqrt{s}\!> \simeq 198.5~\mathrm{GeV}$}. 
A detailed description of the DELPHI detector can be found in reference~\cite{performance}.
In the year 2000 the \cme\ varied from 201.5~GeV to 208.8~GeV,
with an average value of $<\!\!\sqrt{s}\!> \simeq 206~\mathrm{GeV}$.
With the purpose of maximizing the discovery potential, these data were subdivided 
into \cme\ bins that were analysed independently.
During the year 2000 data taking an irrecoverable failure affected one sector of 
the central tracking detector (TPC), corresponding to 1/12 of its acceptance. 
The data recorded under these conditions, approximately 60~pb$^{-1}$, were 
analysed as an independent energy bin. 
The luminosity-weighted mean \cme\ and integrated luminosity for each analysed data  
set are summarised in table~\ref{lum}. 
The last column corresponds to the data taken after the TPC damage.
In the remainder of the text each \cme\ bin will be referred to by the nearest integer value
and the energy bin corresponding to the data taken after the TPC failure as 206$^{*}$.
For the pair-production searches only the data taken in year 2000 were used in 
the analysis. 
In the single-production searches  the 6.9~pb$^{-1}$ collected at $\sqrt{s} \sim 208$~GeV 
were  analysed together with the 207~GeV data.

%%\Small
\begin{table}[hbt]
\begin{center}
\vskip -0.2 cm
\begin{tabular}{||c||c|c|c|c|c|c|c|c|c||}
\hline
\hline
Year             & 1998 & \multicolumn{4}{c|}{1999} & \multicolumn{4}{c|}{2000}    \\
\hline
\hline
%\vskip 0.2 cm
\hspace{-0.3cm} $<\sqrt{s}>$ (GeV) \hspace{-0.3cm} & 
\hspace{-0.25cm} 188.6 \hspace{-0.3cm} & 
\hspace{-0.25cm} 191.6 \hspace{-0.3cm} & 
\hspace{-0.25cm} 195.5 \hspace{-0.3cm} & 
\hspace{-0.25cm} 199.5 \hspace{-0.3cm} & 
\hspace{-0.25cm} 201.6 \hspace{-0.3cm} & 
\hspace{-0.25cm} 204.9 \hspace{-0.3cm} &  
\hspace{-0.25cm} 206.7 \hspace{-0.3cm} & 
\hspace{-0.25cm} 208.2 \hspace{-0.3cm} & 
\hspace{-0.25cm} 206.5 \hspace{-0.3cm} \\
%\vskip 0.2 cm
\hline
$\int{\cal{L}}$ (pb~${}^{-1}$) & 151.8 & 25.1  & 76.0 & 82.6  & 40.1 & 79.9  & 77.1 & 6.9 & 59.2 \\ 
\hline
\hline
\end{tabular}
\end{center}
\vskip -0.45 cm
\caption{Luminosity weighted mean \cme\ and integrated luminosities 
for the analysed data. The last column corresponds to the data taken after the TPC damage.} 
\label{lum}
\end{table}

%%%%%%%%%%%%%%%%%%%%%%%%%%%%%%%%%%%%%%%%%%%%%%%%%%%%%%%%%%%%%%%%%%%%%%%%%%%%%%

Events from Standard Model processes contributing to the background
were generated at each \cme\ using several Monte Carlo programs.
\mbox{$e^+ e^- \rightarrow f \overline{f}  (\gamma)$}  events 
were generated with KK2F~\cite{kk2f} ($f$= quark or muon), KORALZ~\cite{koralz} 
($f$=tau) and \mbox{BHWIDE}~\cite{bhwide} for Bhabha events ($f$=electron). 
Four-fermion final states were produced with WPHACT~\cite{wphact}, while particular 
phase space regions of the 
\mbox{$ {e}^+ {e}^- \rightarrow {e}^+ {e}^- f \overline{f}$}
process, referred to as two-photon interactions, were 
generated using PYTHIA~\cite{pythia} for hadronic final states,
BDKRC~\cite{bdkrc} for $ {e}^+{e}^-\mu^+\mu^-$ and $ {e}^+{e}^-\tau^+\tau^-$
and BDK~\cite{bdk} for ${e}^+{e}^-{e}^+{e}^-$ final states.
${e}^+ {e}^- \rightarrow {e}^+ {e}^- \gamma$ events, 
with one electron (positron) scattered at very small polar angles
while the positron (electron) and photon have large scattering angles, 
yield a final state with only one electron (positron) and one photon detected.  
%%% usually referred to as Compton events, 
Such events, which correspond to a particular region of the Bhabha scattering phase space  
not covered by the BHWIDE simulated sample, 
were generated according to reference~\cite{compton}. 
The process ${e}^+ {e}^- \rightarrow \gamma\gamma(\gamma)$
was simulated using the generator described in reference~\cite{kleiss}.

%%%%%%%%%%%%%%%%%%%%%%%%%%%%%%%%%%%%%%%%%%%%%%%%%%%%%%%%%%%%%%%%%%%%%%%%%%%%%%
Excited-lepton events were simulated to study the distributions of the relevant 
kinematic variables and to compute the selection efficiencies of the analyses.
Single- and pair-production events of all excited-lepton flavours 
were generated according to the differential cross-sections defined in 
reference~\cite{boudjema}. 
Simulated events were produced at the relevant \cmes\ and for several excited 
lepton masses. In the single-production scenario the following masses were 
considered at all \cmes, $m_{\elepton}$ = 100, 125, 150, 170, 180~GeV/$c^2$; 
additional masses were produced up to the kinematic limit, with values depending 
on the \cme\ of the simulated sample (e.g. at $\sqrt{s}= 188.6$~GeV the masses 
185~GeV/$c^2$ and 188~GeV/$c^2$ were also simulated). 
In the pair production the following masses were simulated:
$m_{\elepton}$ = 85, 90, 95, 100 and 103~GeV/$c^2$.  
In addition, for $\ell^* \ell^* \rightarrow \nu \mathrm{W} \nu \mathrm{W}$  
the $m_{\elepton}$ values between 95 and 103~GeV/$c^2$ were taken in 1~GeV/$c^2$ steps.  
  
In all simulations the relation $f=f'$ was assumed. 
However, in the case of the single production of excited electrons, events were 
generated also with \mbox{$f=-f'$}. This allowed to take into account the strong 
dependence of the event kinematics on the relative weights of the couplings.
In the single-production mode, initial-state radiation (ISR) of photons was included 
at the  event generation level, while for the pair-production process it was taken 
into account in the computation of the total cross-section.

All excited-lepton decay modes were included in the single-production 
simulations. For the pair production the following unmixed decays were 
simulated: $\ell^* \ell^* \rightarrow \ell \ell  \gamma \gamma$,
$\nu^* \nu^* \rightarrow \nu \nu  \gamma \gamma$,
$\ell^* \ell^* \rightarrow \nu \mathrm{W} \nu \mathrm{W}$ and 
$\nu^* \nu^* \rightarrow \ell \mathrm{W} \ell \mathrm{W}$.
Finally the decays of \wboson\ and \zboson\  bosons and tau leptons and the 
hadronization/fragmentation in hadronic final states were performed 
using JETSET~7.4~\cite{pythia}.

%%%%%%%%%%%%%%%%%%%%%%%%%%%%%%%%%%%%%%%%%%%%%%%%%%%%%%%%%%%%%%%%%%%%%%%%%%%%%%

The generated signal and background events were passed through the
detailed simulation of the DELPHI detector %%%\ref{delsim} 
and then processed with the same reconstruction and analysis programs 
as the real data~\cite{performance}.
For the data collected after the TPC failure the reconstruction software for 
charged particle tracks was adjusted to make best use of the Silicon Tracker and 
Inner Detector, both placed closer to the beam than the TPC, and the Outer Detector 
and Barrel Rich, placed outside the TPC. 
As a result, the impact of the malfunctioning TPC sector on the determination 
of jet momenta was not large.
A dedicated simulation of the detector conditions during this period was also used.

%%%%%%%%%%%%%%%%%%%%%%%%%%%%%%%%%%%%%%%%%%%%%%%%%%%%%%%%%%%%%%%%%%%%%%%%%%%%%%
\section{Event selection}
\label{sec:selection}

The production and decay of excited leptons would yield  
topologies involving isolated leptons, isolated photons, jets and missing energy,
as detailed in table~\ref{tab:topo}.

In the first step of the analysis, isolated photons and charged leptons  
are searched by constructing double cones centred in the direction of the 
charged particle tracks and the neutral energy deposits, defined as energy 
deposits in the calorimeters not matched to charged particle tracks.
The energy detected inside an inner cone with half opening angle of 5$^\circ$
must be greater than 5~GeV,
while the energy contained between the inner cone and the outer cone must be small
to ensure isolation.
Both the opening angle of the outer cone and the maximal accepted total 
energy contained between the two cones can vary as detailed in reference~\cite{fermiophobic}.

Events are pre-selected by requiring the total energy deposited above
20$^\circ$ in polar angle to be greater than $0.2\sqrt{s}$.  
The events are then classified in different topologies according to their 
multiplicity and to the number of isolated  leptons and photons. 
The  ``low-multiplicity'' events contain at most five well-reconstructed 
tracks while  ``high-multiplicity'' events have more than five such tracks. 
In ``low-multiplicity'' events all particles not identified as isolated photons
are clustered into jets using the Durham algorithm~\cite{durham}. 
This allows to handle the decay products of the tau leptons as low-multiplicity jets.
The jets in the event are obtained by requiring the jet  
resolution variable to be greater than 0.003 for all jet pairings~\cite{delphi2}. 
If the number of jets thus obtained is smaller than the number of isolated 
leptons previously found, the algorithm is applied once more requiring the number 
of jets to be equal to the number of isolated leptons.
For the ``high-multiplicity'' events no dedicated strategy for the taus was followed.  
In this case a tau is reconstructed only if its decay products 
(charged leptons or low-multiplicity jets) fulfil the double cone isolation criteria. 
Electron, photon and muon identification are based on the standard 
DELPHI algorithms described in reference~\cite{performance}.
Topology-dependent criteria are finally applied, as detailed in the following 
subsections, and, whenever possible, the flavour of the final-state 
leptons is used to tag the flavour of the excited lepton. 

\subsection{Topologies with only photons}
\label{sec:photons}

Final-state topologies consisting of photons only could arise from the
production of excited neutrinos decaying to a SM neutrino and a photon. 
For these topologies the analyses presented in references~\cite{fermiophobic}
and \cite{single-photon} were used.

In the search for single production of excited neutrinos,  
events consisting of only one photon in DELPHI (single-photon events) are considered. 
For the single-photon preselection the results from   
reference~\cite{single-photon} were used.
The background from SM processes giving single-photon events 
is mainly due to the process 
${e}^+ {e}^- \rightarrow \zboson \gamma \rightarrow \nu \overline{\nu} \gamma$,
where the final-state photon is emitted predominantly at small polar angles.
Candidate events must have a photon with polar angle 
$\theta_\gamma > 45^\circ$ and with energy $E_\gamma > 0.45 \sqrt{s}$.
In the $\nu \nu^*$ search the data from year 2000 was grouped in two bins, 
corresponding to $\sqrt{s} < 207$~GeV (including the 206$^{*}$ data) and $\sqrt{s} \ge 207$~GeV. 

For the $\nu^* \nu^*$ search events with two photons were selected. The selection of 
events with two photons and missing energy described in 
reference~\cite{fermiophobic} was followed, except for  the kinematic fit 
imposing the \zboson\ mass on the invisible system and the requirement on the 
missing mass.

\subsection{Topologies with leptons and photons}
\label{sec:lepsel}

Topologies with isolated leptons and photons are expected whenever
the excited charged leptons decay by photon emission.

For the single-production search the topologies $\ell\ell\gamma$ and
$\ell\gamma$ were considered as shown in table~\ref{tab:topo}.
The $\ell\gamma$ topology becomes dominant for all flavours when the excited-lepton mass 
is close to the \cme. The spectator SM lepton has then too small an energy to be identified 
as an isolated particle. The  $\ell\gamma$ topology is also crucial for the single $e^*$
search when $t$-channel production dominates, in particular for the $f=f'$ scenario. 
The SM spectator electron is then scattered at small polar
angles, remaining undetected.

Different preselection criteria were applied, according to the event classification 
in each of the topologies and taking into account the relevant background 
processes and the specific kinematics of the signal events. 

Only events with at least one photon with energy $E_\gamma > 0.05\sqrt{s}$  are 
considered; for the lepton momentum ($p_\ell$) it is required $p_\ell > 0.05\sqrt{s}$ 
in the topologies with only one lepton. 
In the $\ell\ell\gamma$ topology the sum of the two lepton momenta must
be greater than $0.1\sqrt{s}$. The sum of the lepton and photon 
energies ($p_{\ell_1} +p_{\ell_2} +  E_\gamma$) should in principle 
be of the order of the \cme;
however, since in $\tau \tau^* \rightarrow \tau \tau \gamma$ events a
fraction of the energy is carried by neutrinos,
it is required $p_{\ell_1} +p_{\ell_2} +  E_\gamma > 0.4\sqrt{s}$.  
In the $e^*$ and $\mu^*$ searches, 
the momentum of the most energetic lepton must be greater than $0.05\sqrt{s}$. 
For events in the  $\ell\gamma$ topology it is required that $E_\gamma > 0.1\sqrt{s}$, 
$p_\ell > 0.1\sqrt{s}$ and $E_\gamma + p_\ell > 0.4\sqrt{s}$.

\begin{figure}[p]
\begin{center}
\vspace{-1.5cm}
\mbox{\epsfig{file=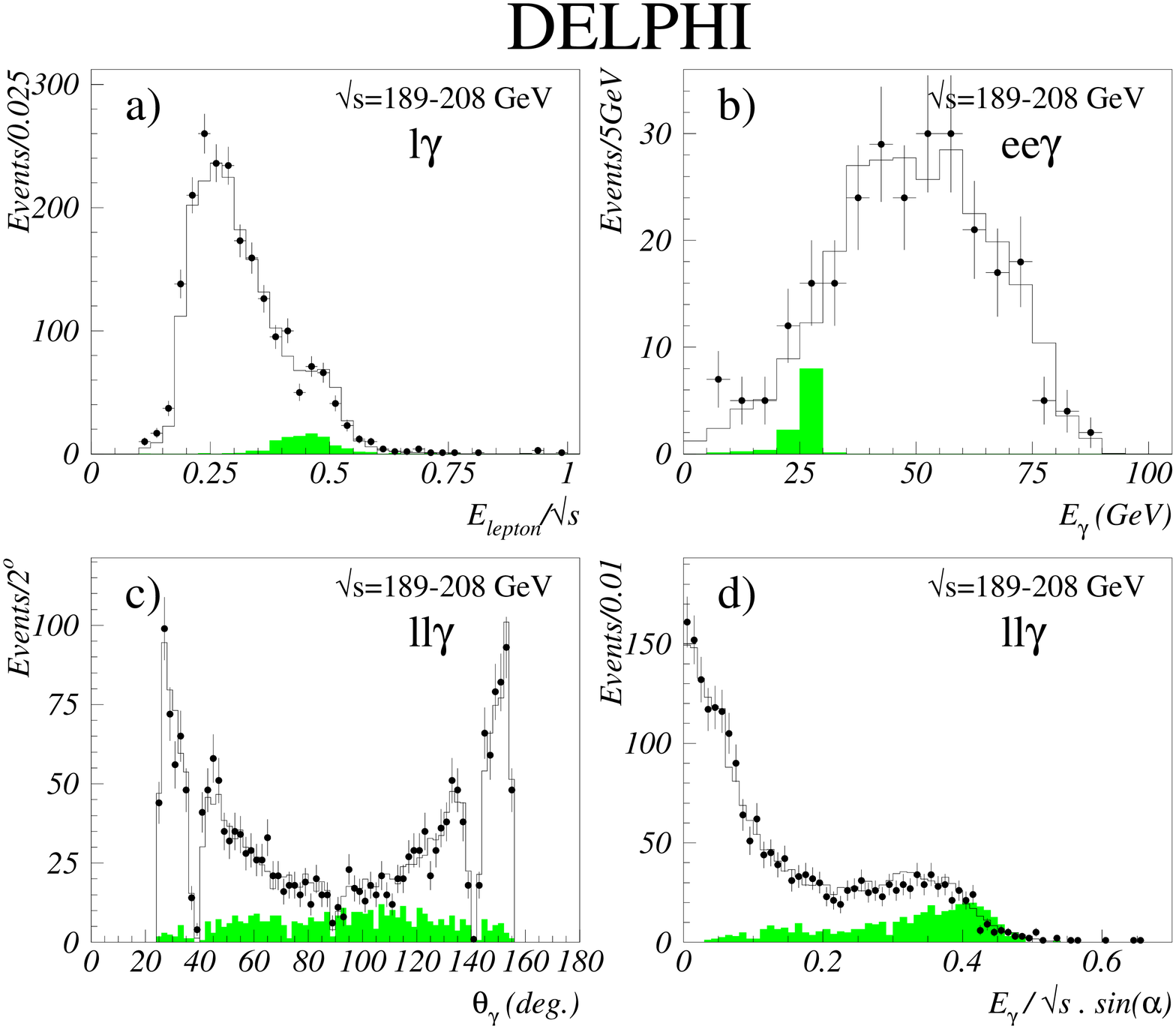,width=0.975\textwidth}}
\vspace{-0.2cm}
\caption[]{Topologies with leptons and photons after the preselection cuts: 
{\bf (a)} lepton energy in the $\ell\gamma$ topology; 
{\bf (b)} energy of the least energetic photon in the e$^*$ search in the final-state 
topology with one electron and two reconstructed photons;
{\bf (c)} photon polar angle and   
{\bf (d)} $E_\gamma/\sqrt{s} \cdot \sin \alpha $, where $\alpha$ is the
minimum angle between the photon direction and the two lepton directions, 
in the $\ell\ell\gamma$ topology.
The dots show the data and the white histograms show the SM simulation.
The shaded histograms show the expected distributions for a 
$m_{\elepton}=175$~GeV/$c^2$
excited lepton at $\sqrt{s}=206$~GeV, using an arbitrary normalization.}
\label{fig:lept}
\end{center}
\end{figure}

Figure~\ref{fig:lept} shows the distribution, after this preselection, of relevant 
kinematic variables for each of the topologies considered. 

The event selection was further tightened as follows.
In the case of the $\ell\gamma$ topology the dominant background arise from Bhabha 
scattering events with one electron lost at low polar angle or identified as a photon. 
Requiring the presence of an energetic photon in the central region of the detector is 
the main rejection criterium. 
For the $e^*$ and $\tau^*$ searches in this topology  it is thus required that 
$\theta_\gamma > 42^\circ$.

The main background for the $\ell\ell\gamma$ topology is due to radiative 
Bhabha scattering events.
Initial-state radiation events, where the photon is mainly emitted at small polar angle,
are reduced by requiring $\theta_\gamma > 42^\circ$ in the  $e^*$ and $\tau^*$ searches.
The background from final-state radiation consists of low-energy photons emitted at 
small angle to the direction of the final-state leptons. 
It is thus required  $E_\gamma \cdot \sin \alpha > 0.08\sqrt{s}$, where $E_\gamma$ is
the photon energy and $\alpha$ is the angle between the photon direction and the
direction of the nearest particle.

The final-state topology consisting of one electron and two photons, e$\gamma\gamma$,   
was additionaly considered in the $e^*$ search. 
In the $t$-channel-dominated $e^*$ production mode the 
spectator electron scattered in the forward direction
could be detected by the low-angle calorimeter of DELPHI, below the geometrical acceptance 
of the tracking detectors, thus being reconstructed as a photon.
For the selection of events in the e$\gamma\gamma$ topology only one 
photon with $\theta_\gamma < 9^\circ$ can be present; the other photon must be detected 
above  $25^\circ$;
the sum of the lepton momentum and photon energies is above $0.8\sqrt{s}$. 
The main background for this topology is due to radiative Bhabha scattering events, with 
one low angle electron being identified as a photon. This background is mostly 
irreducible.   
Some reduction of the background is achieved by requiring that 
$p_\ell+E_{\gamma_1} > 0.4\sqrt{s}$,  where
$E_{\gamma_1}$ is the energy of the low polar-angle photon.

For the pair-production searches the $\ell\ell\gamma\gamma$ topology was considered. 
The expected background is rather low and simpler cuts were applied. 
Both leptons must have momentum above 10~GeV/$c$. 
Events are kept as candidates if a lepton-photon pairing exists for which 
the difference between the  invariant masses of the two lepton-photon pairs is 
smaller than 15~GeV/$c^2$ (20~GeV/$c^2$) in the $e^*$ and $\mu^*$ ($\tau^*$) searches.

The excited-lepton mass can be reconstructed by computing the lepton-photon invariant mass.
In the $\ell\gamma\gamma$ topology the photon expected from the decay of
an excited lepton is the one detected at high polar angle, while in the $\ell\ell\gamma$ topology both possible lepton-photon pairings are considered.
The invariant-mass resolution improves by rescaling the measured energies and momenta. 
This is done imposing energy-momentum conservation and using just the polar and azimuthal 
angles, which are well measured in the detector.
Resolution of $\pm 1$~mrad in $\theta$ and $\pm 1.7$~mrad in $\phi$ are obtained for 
high energy photons and of about
$\pm 1$~mrad or better in $\theta$ and $\phi$ are obtained for high momentum charged 
particle tracks, in the central part of the DELPHI detector~\cite{performance}.  
%%%
In order to take into account the energy lost through initial-state radiation,
the rescaling is also applied assuming the  presence of an additional photon along 
the beam direction.  
This procedure accounts also for the case when the spectator
electron is lost in the beam pipe.
The compatibility of the rescaled and the measured values is
quantified through the $\chi^2$ parameter~\cite{delphi1,delphi2}. 
%%% defined as:
%%%
%%% $$ 
%%%  \chi^2~=~\frac{1}{n} \sum_{i=1,n}
%%%  \left(\frac{p_i^{calc}-p_i^{meas}}{\sigma_i}\right)^2
%%% $$
%%%
%%%\noindent
%%%where $n$ is the number of particles, 
%%%$p_i^{meas}$ are the measured momenta or energies and
%%%$p_i^{calc}$ are the values calculated from the kinematic constraint;  
%%%$\sigma_i$, the quadratic sum of the errors on $p_i^{calc}$ and $p_i^{meas}$, 
%%%are defined in reference~\cite{delphi1}.
The $\chi^2$ was computed separately for charged particles ($\chi^2_{charged}$) and photons 
($\chi^2_{photons}$). 
The result from the rescaling assuming an additional particle along the beam direction 
was retained whenever  it  yielded a  smaller total $\chi^2$. 
In any case only events with 
$\min(\chi^2_{\mathrm{charged}},\chi^2_{\mathrm{photons}})<5$ are retained. 
The resolution on the lepton-photon invariant mass, 
after applying the kinematic constraints, is in the range $0.2-0.6$~GeV/$c^2$
($1.5-2.0$~GeV/$c^2$) for electrons and muons (taus). 

Finally, the flavour of the final-state leptons is used to select the candidate events.
In the $e^*$ search all leptons must be identified as electrons.
The search for $\mu^*$ requires that the most energetic lepton is identified 
as a muon and no particle is identified as an electron.
In the $\tau^*$ search  no lepton flavour identification is applied; 
instead, a difference between the measured and rescaled momenta of the final-state 
leptons, characteristic of the presence of neutrinos from tau decays, 
is required by imposing $\chi^2_{charged}>5$  in the single-production 
and $\chi^2_{charged}>10$ in the pair-production searches. 

\subsection{Topologies with jets and leptons}
\label{sec:hadsel}

Final-state topologies with jets and isolated leptons were considered in the search for 
excited leptons decaying to \wboson\ or \zboson,   
$\ell^* \rightarrow \nu \mathrm{W}$, $\ell^* \rightarrow \ell \zboson$,
$\nu^* \rightarrow \ell \mathrm{W}$ and $\nu^* \rightarrow \nu \zboson $.
Due to the presence of neutrinos, some of the topologies are additionaly characterized 
by missing energy.
In the single-production searches only the final states arising from the hadronic decays 
of the \wboson\ and \zboson\ bosons were addressed. In the pair-production searches   
only the fully hadronic or semileptonic decays of the  \wpair\ pairs were addressed.
The  ``high-multiplicity'' events were thus considered in these analyses.

\subsubsection{Single-production analysis}
\label{sec:single}

The topologies considered in the single-production search are 
$jj\!\!\eslash$, $jj\ell\!\!\eslash$,  $jj\ell$ and $jj\ell\ell$.
All particles in the event, excluding the isolated leptons, are clustered 
into jets using the Durham algorithm. Two-jet events are  
selected by requiring the Durham jet resolution variable for the transition from  
three to two jets, $y_{23}$, to be lower than 0.06 and from 
two to one jet, $y_{12}$, to be greater than 0.01.
The polar angle of isolated leptons must be above $25^\circ$.

The searches for pair production of excited leptons already excluded 
\elepton\ masses smaller than the mass of the \zboson\ boson for all excited 
lepton flavours. 
The gauge bosons are thus expected to be produced on-shell 
and the invariant mass of the two jets 
($M_{jj}$) should be compatible with a \wboson\ or \zboson\ boson. 
The loose condition $40 < M_{jj} < 120~{\mathrm{GeV}}/c^2 $ is thus applied 
in all topologies.  
In addition, since the gauge bosons originating from excited-lepton decays are not at rest, 
the two jets are also expected to be acoplanar. This characteristic is quantified by    
the jet-jet acoplanarity, $A^{jj}_{cop}$,  defined as 180$^\circ-\Phi$, where
$\Phi$ is the angle between the projections of the jet momenta in the plane 
perpendicular to the beam.

For events in the $jj$ topology the main background comes from 
$e^+ e^- \rightarrow q \overline{q}(\gamma)$ events, where the photon is emitted 
at a very small polar angle or is soft, and thus remains undetected. 
Since the transverse momentum of the photon is always very small 
(typically $< 2$~GeV/$c$ at $\sqrt{s}=200$~GeV), 
this process results in two jets with small acoplanarity.
Candidate events are required to have $A^{jj}_{cop}>25^\circ$ and
the polar angle of both jets larger than 20$^\circ$.

A looser acoplanarity cut, $A^{jj}_{cop}>10^\circ$, is applied to events in the  
$jj\ell$ topology.
The background is mainly due to \wpair\ production, with one boson decaying to quarks 
and the other to a charged lepton and a neutrino (\wboson\ semileptonic decays). 
The quantity $\xi = {\mathrm{Q}}_{\mathrm{W}}\cdot cos \theta_{\mathrm{W}}$,
$\mathrm{Q}_{\mathrm{W}}$ and $\theta_{\mathrm{W}}$ being the boson charge and 
polar angle respectively, was used to reduce the \wpair\ background. 
The \wboson\ bosons in background events are produced 
preferentially in the forward direction and $\xi$ is peaked towards -1.
In \wboson\ semileptonic decays $\mathrm{Q}_{\mathrm{W}}$ is given by the lepton charge 
and $\theta_{\mathrm{W}}$ is estimated, neglecting radiation 
effects, from the jet directions.
%%% as $180^\circ-\theta_{\overrightarrow{J}}$, 
%%% where $ \theta_{\overrightarrow{J}} $  is the polar angle of the sum of the momenta of the
%%% two jets after the kinematic fit described below. 
For events in the $jj\ell$ topology with the lepton charge unambiguously determined, 
about 90\% of the events, it is required $\xi > -0.6$, while no condition is 
applied to the remaining events.

In the $jj\ell\ell$ topology the angle between the two lepton directions or between any
of the leptons and the jet directions must be greater than $10^\circ$. 
No acoplanarity cut is applied.

\begin{figure}[p]
\begin{center}
\vspace{-1.5cm}
\mbox{\epsfig{file=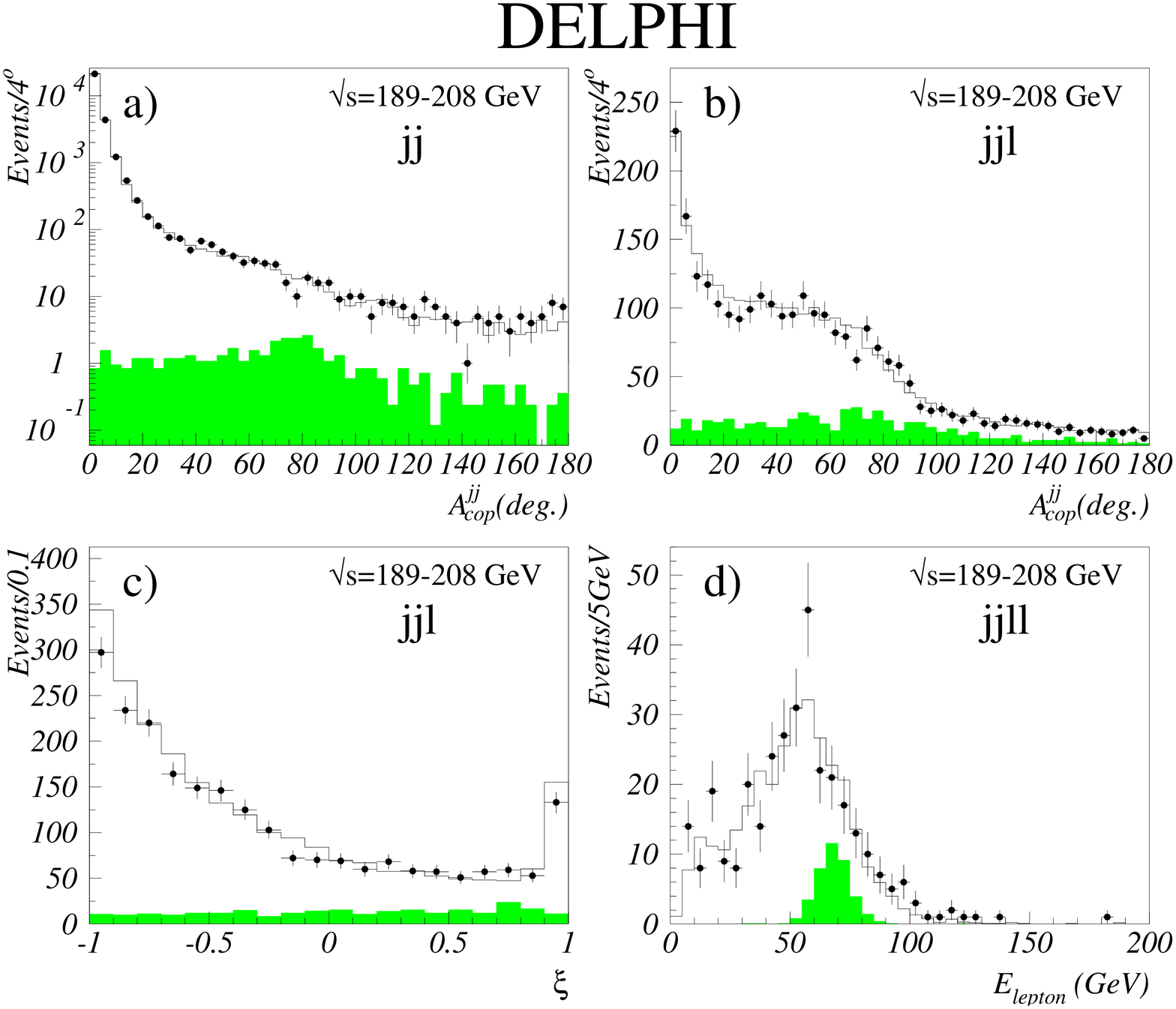,width=0.975\textwidth}}
%\vspace{0.5cm}
%\vspace{-1cm}
\caption[]{Topologies with jets and leptons after the preselection cuts:
{\bf (a)} jet-jet acoplanarity in the $jj$     topology;
{\bf (b)} jet-jet acoplanarity in the $jj\ell$ topology;
{\bf (c)} variable $\xi = \mathrm{Q_W} \cdot \cos \theta_{\mathrm{W}}$ 
in the $jj\ell$ topology, for events with the lepton charge unambiguously 
determined;  
{\bf (d)} energy of the most energetic lepton in  $jj\ell\ell$ events.
The dots show the data and the white histograms show the SM simulation.
The shaded histograms show the expected distributions for a 
$m_{\elepton}=175$~GeV/$c^2$ excited lepton at $\sqrt{s}=206$~GeV, 
using an arbitrary normalization.}
\label{fig:had}

\end{center}
\end{figure}

Figure~\ref{fig:had} shows the distributions for the jet-jet acoplanarity, 
the variable $\xi$ for events with the lepton charge unambiguously determined 
and the energy of the most energetic lepton,   
for the  $jj$,  $jj\ell$ and  $jj\ell\ell$ topologies, after the preselection cuts.
%The different topology dependent selection criteria are described in the following 
%paragraphs.

In order to improve the estimation of the jet momenta and energies 
a kinematic fit~\cite{fit5c} is applied to the selected events.
Events in the  $jj$ and $jj\ell$ topologies can arise from excited-lepton decays 
mediated by a \wboson\ or \zboson\ boson  and  thus the invariant mass of 
the jet-jet system is constrained to be either $m_W$ or $m_Z$, 
depending on the search channel. 
For events in the $jj\ell\ell$ topology only the $m_Z$ constraint is used.
In all cases, only the events compatible with the expected decay mode are
retained by requiring that the kinematic fit yielded a $\chi^2$ per degree of 
freedom lower than 5.

The excited-lepton mass can be estimated in several of the topologies considered. 
The relevant variables are the 
jet-jet-lepton  (\mbox{$\nu \nu^* \rightarrow \nu \ell \wboson $})
and  
jet-jet-neutrino (\mbox{$\ell \ell^* \rightarrow \ell \nu \wboson $}) invariant masses, 
and the recoil mass of isolated leptons 
(\mbox{$\ell \ell^* \rightarrow \ell \nu \wboson $}, 
\mbox{$\ell \ell^* \rightarrow \ell \ell \zboson $}).
The neutrino four momenta ($P_\nu$) was reconstructed, from the total energy 
($E$) and momenta ($\overrightarrow{p}$) of all 
measured final-state particles, as 
$P_\nu = (\overrightarrow{0},\sqrt{s}) - (\overrightarrow{p},E)$. 
%%%%%%%%%%%%%%%%%%%%%%%%%%%%%%%%%%%%%%%%%%%%%%%%%%%%%%%%%%%%%%%%%%%%%%%%%%%%%%%%%
The resolution on the jet-jet-neutrino invariant mass varies between 
1~GeV/$c^2$ and 5~GeV/$c^2$.
In the  $\tau\tau^* \rightarrow \tau \nu \wboson $ channel, the $\tau^*$ mass 
is reconstructed only for signal masses \mbox{$m_{\tau^*} > 0.9\sqrt{s}$} and
is obtained from the recoil mass of the spectator lepton;
the resolution ranges between 3~GeV/$c^2$ and 6~GeV/$c^2$. 
The resolution on the jet-jet-lepton invariant mass is about 2~GeV/$c^2$ for 
\mbox{$m_{\nu^*}=100~{\mathrm{GeV}}/c^2$}, increasing to about 10~GeV/$c^2$ for
\mbox{$m_{\nu^*}=200~{\mathrm{GeV}}/c^2$}; 
no mass reconstruction was attempted in the $\nu^*_{\tau}$ channel.

At the last step of the analysis, the various production and decay modes
within the same topology are treated differently.
In the searches for excited leptons of the first and second generations, 
the flavour of the final-state isolated leptons must match the excited-lepton flavour.
In the searches for $\tau^*$ and $\nu_{\tau}^*$ in the topologies with isolated leptons
no selection is applied based on the  flavour of the final-state leptons. 
Instead, due to the presence of  neutrinos in the tau decay products, the final-state 
isolated leptons are expected to have relatively small energy. Therefore a cut on the
energy of the final-state leptons is used  as follows: in  
the $\tau^* \rightarrow \tau \zboson$ search, the energy of the isolated lepton  
must be lower than $0.3 \sqrt{s}$ for events in the $jj\ell$ topology, while 
in the $jj\ell\ell$ topology at least one lepton with energy 
smaller than $0.2 \sqrt{s}$ must be present; 
in the $\nu_{\tau}^* \rightarrow \tau \wboson$ search the lepton energy must be 
smaller than $0.2\sqrt{s}$.

\subsubsection{Pair-production analysis}

\underline{$\nu^*\nu^*$ search}
\vspace{0.25cm}

In the search for pair production of neutral excited leptons decaying to \wboson\ bosons
($\nu^* \nu^* \rightarrow \ell \mathrm{W} \ell \mathrm{W}$)  
only the fully hadronic and semileptonic decays of \wpair\  pairs were taken into account.
The final-state topologies are formed by the \wpair\ decay products
(two jets and one lepton or four jets) and by two additional charged leptons, 
yielding a rather clear signature.
%% The background is mainly due to \mbox{\zboson\zboson} events.

Multijet events are selected by requiring $y_{12}>0.03$ and only
events with at least two isolated leptons are kept.
If exactly two isolated leptons are found it is further required 
that $y_{23}>0.01$. 

In the search for $\nu^*_e$ and $\nu^*_{\mu}$ , the final state
must contain at least two charged leptons of the corresponding flavour.
For the $\nu^*_{\tau}$  search it is required  $\eslash > 0.1 \sqrt{s}$.

\vspace{0.5cm}
\underline{$\ell^*\ell^*$ search}
\vspace{0.25cm}

The final-state topologies considered in the search for pair production of charged 
excited leptons decaying to \wboson\ bosons
($\ell^* \ell^* \rightarrow \nu \mathrm{W} \nu \mathrm{W}$)  
were four jets or two jets and one lepton. They  result respectively 
from the fully hadronic or semileptonic decays of the \wpair\ pair.

Contrary to the $\nu^*\nu^*$ search described above, the two additional particles 
in the final state are now neutrinos, giving missing energy.
Signal events have thus a signature very similar to the  \wpair\ background events.
A discriminant analysis was used 
in the $\ell^* \ell^* \rightarrow \nu \mathrm{W} \nu \mathrm{W}$ search,
in order to boost the small differences between the signal and background kinematics
After the event preselection  a signal likelihood, ${\cal L}_{S}$,
and a background likelihood, ${\cal L}_{B}$, are constructed as the product of
probability density functions (PDFs) of relevant kinematic
variables, as described below. 
The discriminant variable is defined as ${\cal L}_{S}/{\cal L}_{B}$.

The semileptonic and the fully hadronic cases were treated separately. 
In the semileptonic analysis only events with no isolated photons and at least one 
isolated  lepton are considered. The remaining particles in the event are
clustered into jets. Two-jet events are selected by requiring 
%%% that the Durham resolution variables satisfy the criteria  
$y_{23}<0.06$ and $y_{12}>0.01$.
The background from $e^+ e^- \rightarrow q \overline{q}(\gamma)$ events
is reduced by requiring the polar angle of the direction of the 
missing momentum to be above $20^\circ$. 
The minimum transverse momentum of the lepton with respect to any of the jets
must be greater than 10~GeV/$c$; the lepton polar angle is
required to be above 20$^\circ$ for muons and above 40$^\circ$ for electrons. 
The following variables are then used to build the discriminant variable: 
\begin{itemize}
\item
the missing energy of the event;
\item
the angle between the two jets;
\item
the energy of the lepton;
\item
the angle between the lepton and missing momentum directions; 
\item
the $\mathrm{Q}_{\ell} \cdot \cos \theta_{\ell}$ variable, where 
$\mathrm{Q}_{\ell}$ and $\theta_{\ell}$ are the charge and polar angle of the lepton.
\end{itemize}

In the fully hadronic analysis it is required that no isolated photons or leptons 
are found.  Four-jet events are selected by requiring
$y_{34}>0.003$ and $y_{23}>0.03$. 
The jets are assigned to each of the \wboson\ bosons by choosing the pairing
that minimizes the sum of the squares of the differences between the jet-jet 
invariant masses and the \wboson\  mass.     
A fit imposing energy-momentum 
conservation and constraining the invariant mass of the two jet pairs to the  \wboson\ 
mass is performed. 
The following quantities are then used to build the discriminant variable:
\begin{itemize}
\item
the missing energy of the event;
\item
the angle between the directions of the two jets of each pair;
\item
the angle between the two reconstructed W bosons.
\end{itemize}

\begin{figure}[p]
\begin{center}
\vspace{-1.5cm}
\mbox{\epsfig{file=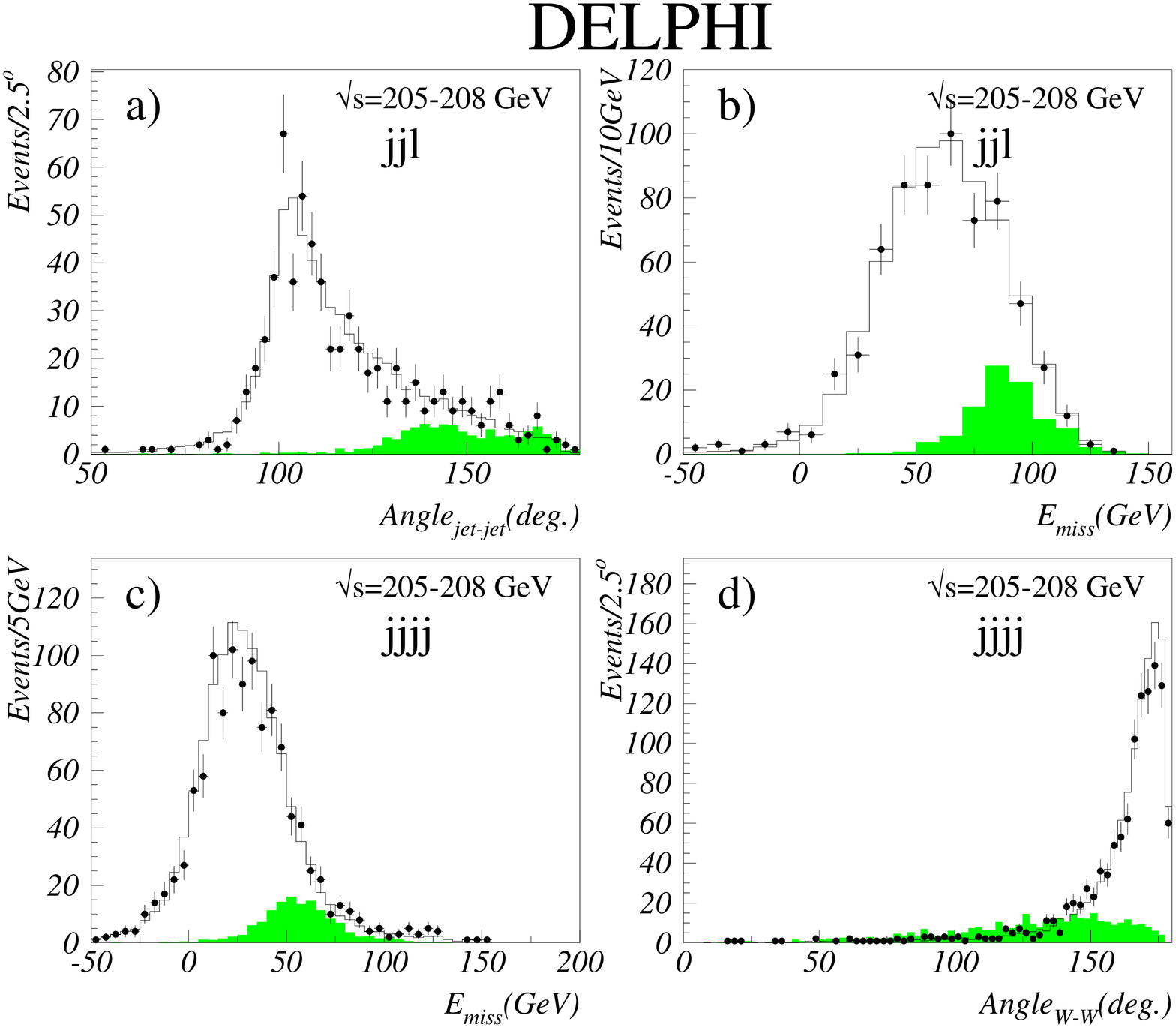,width=0.975\textwidth}}

\vspace{-0.1cm}
\caption[]{$\ell^*\ell^*$ search at the preselection level: 
{\bf(a)} angle between the two jets  and {\bf(b)} missing energy  
 in the semileptonic channel; 
 {\bf(c)} missing energy and {\bf(d)} angle between the two reconstructed $W$s  in the
fully hadronic channel. 
The dots show the data and the white histograms show the expected 
SM background. 
The shaded histograms show the expected signal distributions at $\sqrt{s}=206$~GeV
with m$_{\elepton}$=100~GeV/$c^2$, using an arbitrary normalization.}
\label{fig:double1}

\end{center}
\end{figure}

The distributions of some of the variables used to build 
${\cal L}_{S}/{\cal L}_{B}$ are shown in figure~\ref{fig:double1}. 
A good agreement with the SM predictions is observed.
It should be noted that due to the finite resolution in the measurement of
the charged particle tracks and calorimeter energy deposits, the missing energy 
of the event \mbox{$\eslash =  \sqrt{s} - E_{\rm vis}$}, with $E_{\rm vis}$ the total 
measured energy in the event) may fluctuate to negative values.

\begin{figure}[H]

\begin{center}
\vspace{-0.5cm}
\mbox{\epsfig{file=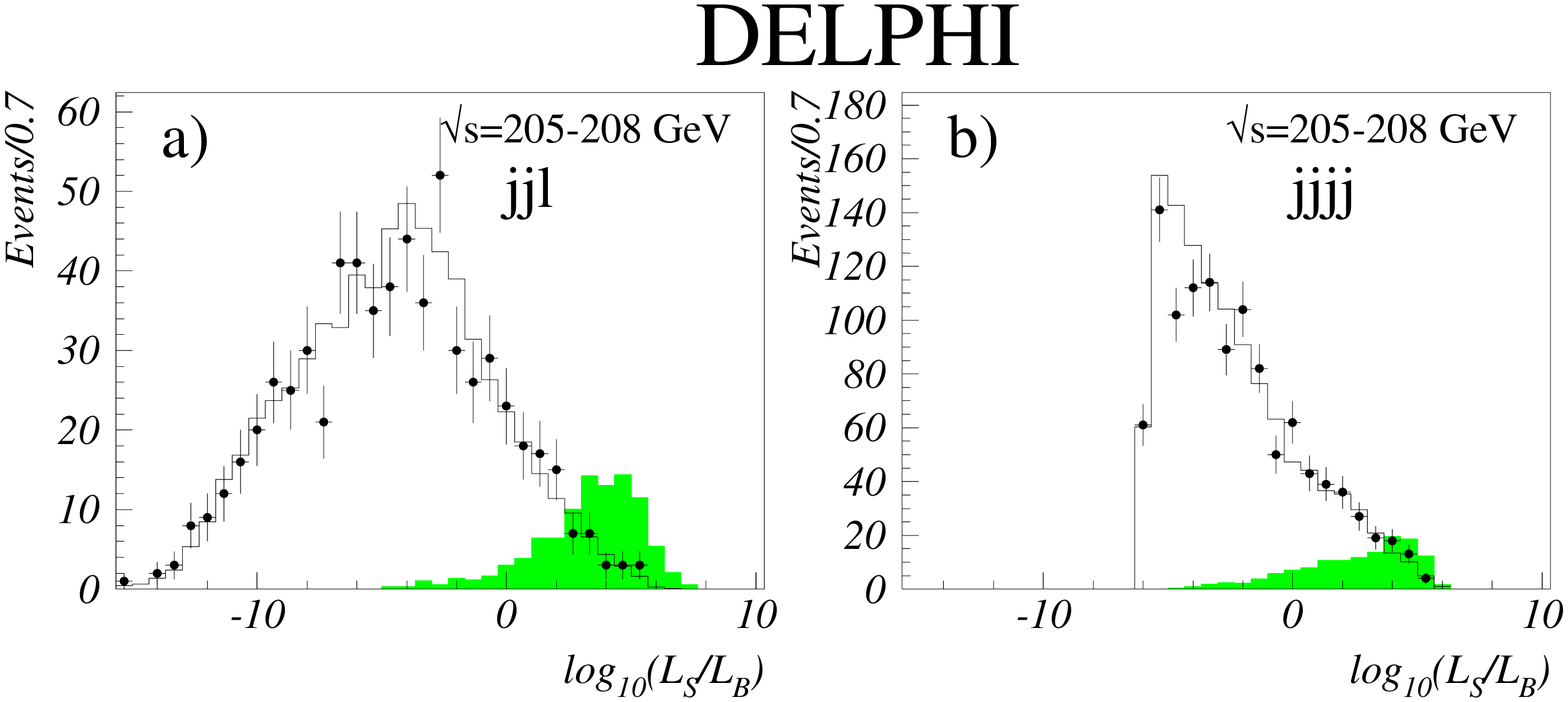,width=0.95\textwidth}}

\vspace{-2mm}
\caption[]{$\ell^*\ell^*$ search: discriminant variables in the 
{\bf (a)} semileptonic   and {\bf (b)} fully hadronic  
channels. The dots are the data and the white  histograms show 
the SM background expectation. 
The shaded histograms are the expected distributions for a 
m$_{\elepton}$=100~GeV/$c^2$ signal at $\sqrt{s}=206$~GeV, 
using an arbitrary normalization.}
\label{fig:double2}

\end{center}
\end{figure}

In figure~\ref{fig:double2} are shown the distributions of ${\cal L}_{S}/{\cal L}_{B}$ 
for the semileptonic and the hadronic final states.

\section{Results}
\label{sec:results}

No evidence for the production of excited leptons was observed in any of the final 
states considered.
The number of candidate events found  at the various \cmes, 
together with the expected background from SM processes, are summarized in 
tables~\ref{tab:single} and \ref{tab:pair},
for the different excited-lepton flavours and decay modes. The total numbers
are summarized in table~\ref{tab:all}.
These numbers are obtained by adding the results from the different exclusive final
state topologies considered in each decay (as listed in table~\ref{tab:topo}). 
The relevance of each topology depends on the decay branching ratios, which are 
a function of the excited-lepton mass and of the coupling parameters.
In many cases there are candidate events common to the different excited-lepton searches 
(e.g. the events selected in the $jj$ topology are candidates in 
all $\ell^*\rightarrow \nu W$ searches, independent of the $\ell^*$ flavour), 
but in the search for a given flavour there are no common candidates selected 
in final states originating from different decay modes.

The signal selection efficiencies at $\sqrt{s}=206$~GeV 
are given in table \ref{tab:effi}, for specific values of $m_{\elepton}$.
In most of the channels the dependence of the efficiency on the mass is weak, 
as we benefited from the combination of results from several final-state topologies 
which are sensitive to different mass regions. 

\begin{figure}[p]
\begin{center}
\vspace{-1.2cm}
\mbox{\epsfig{file=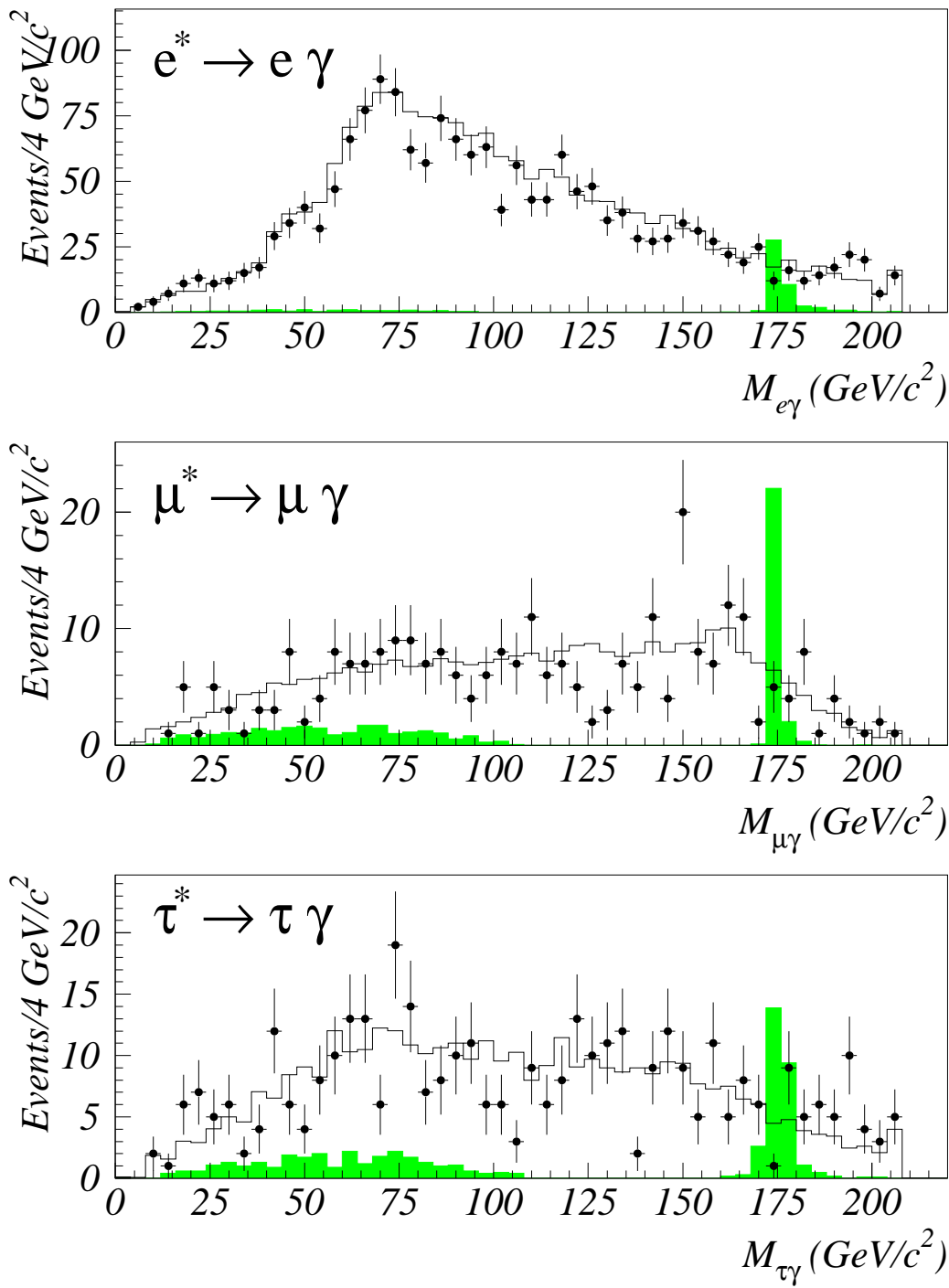,width=0.95\textwidth}}
\vspace{-1.2cm}
\caption{Lepton-photon invariant masses for the selected candidates in the 
$\ell^* \rightarrow \ell \gamma$ channels. 
Events selected in all relevant final-state topologies at all centre-of-mass energies were added.
For events from the $\ell \ell \gamma$ final-state topology
the two possible $\ell \gamma$ combinations are shown.  
The dots show the data and the white histograms show the expected SM background. 
The shaded histograms show the expected signal distributions at $\sqrt{s}=206$~GeV
with m$_{\ell^*}$=175~GeV/$c^2$, using an arbitrary normalization.}
\label{fig:lep-masses}
\end{center}
\end{figure}

\begin{figure}[p]
\begin{center}
\vspace{-1.0cm}
\mbox{\epsfig{file=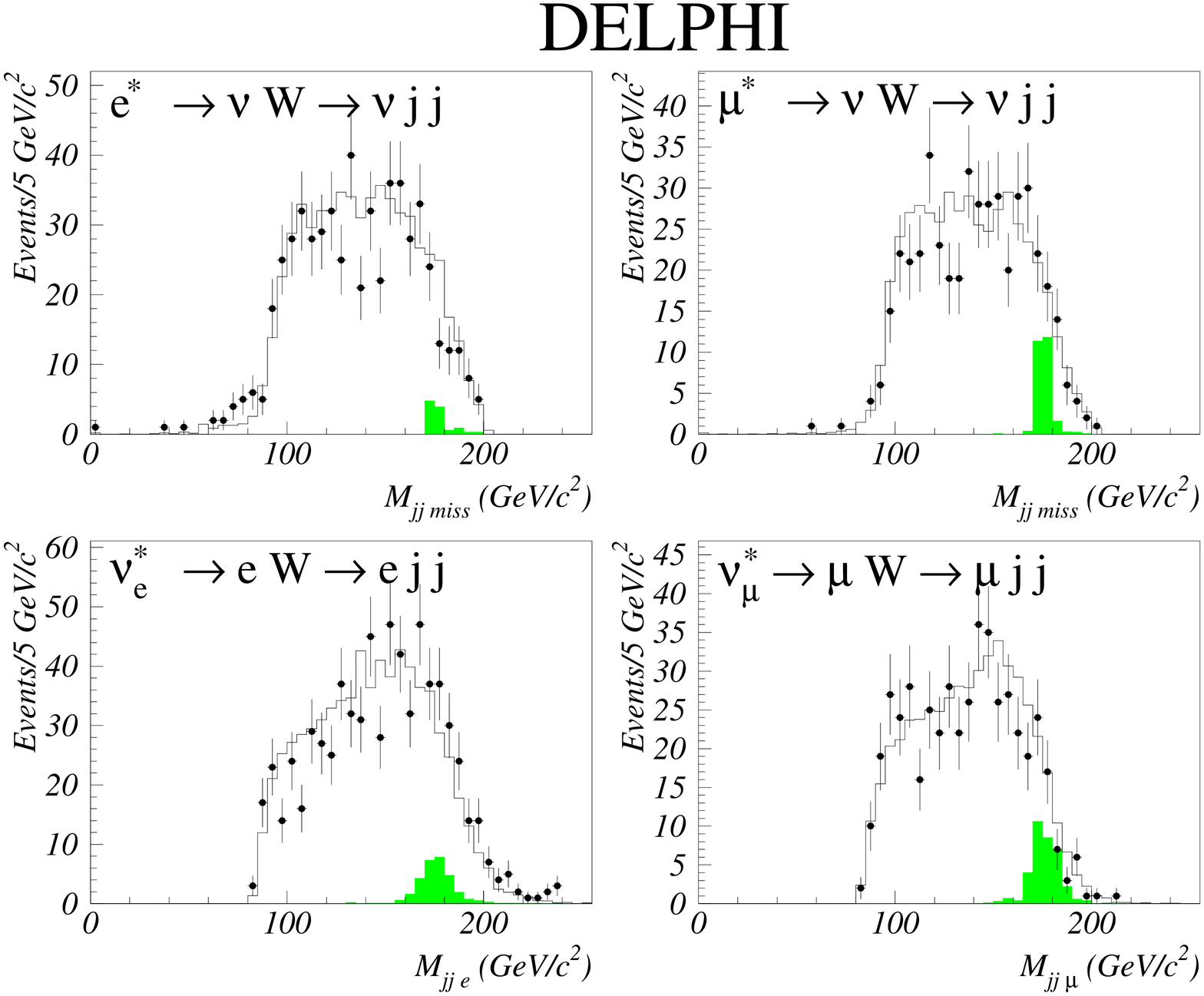,width=0.95\textwidth}}
\vspace{-1.cm}
\caption{Invariant mass distributions for the selected candidates in the  
$\ell^* \rightarrow \nu W$ (top) and $\nu^* \rightarrow \ell W$ (bottom) channels.
Events selected in the $jj\ell$ topology at all centre-of-mass energies were added.
The dots show the data and the white histograms show the expected SM background. 
The shaded histograms show the expected signal distributions at $\sqrt{s}=206$~GeV
with m$_{\elepton}$=175~GeV/$c^2$, using an arbitrary normalization.}
\label{fig:had-masses}
\end{center}
\end{figure}

Figures~\ref{fig:lep-masses} and \ref{fig:had-masses} show the invariant mass 
distributions for the candidates selected in the various single-production searches, 
obtained by adding the data from all the analysed  \cmes\ and topologies.

\subsection{Systematic uncertainties}

Systematic uncertainties affect both the background and the 
signal efficiency estimations. 

Theoretical errors on the computed cross-sections
translate into uncertainties on the expected number of background events, 
%%% The overall error on the number of background events resulting from 
%%% the systematic errors on the cross-section of processes contributing 
%%%significantly to the background is 
typically less than 2\%~\cite{LEP2MC}. 

At the event generator level, the simulated distributions of the kinematic variables 
used in the event selection may not match the distributions for the data,
due either to an imperfect description of the detector or of the 
background processes at the event generation level. 
Possible effects on the selected number of simulated events were estimated 
by studying the change in the ratio between the number of selected events in the 
data and simulated background when varying each selection cut around the nominal value.  
The most relevant variables used in the event selection were changed as follows:  
the photon polar angle was varied by
$\pm 5^\circ$; the cut in the lepton energy was changed by $\pm 5$~GeV;
the jet-jet acoplanarity cut was varied by $\pm 5^\circ$;
the limits in the jet-jet invariant mass window were changed by 
$\pm 10$~GeV.
In each topology the contributions from the different selection cuts were added 
in quadrature. The total systematic uncertainty  from the event selection cuts 
ranged between   5\% and 8\%, depending on the topology considered. 
These were taken as an estimate of the contribution from the analysis 
cuts to the systematic error on the background expectation. 
The systematic uncertainties on the signal selection 
efficiencies due to the selection cuts were assumed to be equal to and 
fully correlated with the errors estimated above for the background. 

The statistical errors on the background and signal efficiencies, due to the limited
Monte Carlo statistics,  were taken as uncorrelated systematic uncertainties.
 
%In the leptonic topologies, the cut on the photon polar angle was varied by 
%$\pm 5^\circ$ giving a change of about 3\% in the Data/MC ratio;
%the cut in the minimum energy of the lepton gives no significant contribution
%to the systematic error. 
%In the $\ell\ell \gamma$ topology the cut on 
%$\mathrm{E}_{\gamma}\times \sin{\alpha^{\mathrm{iso}}_{\gamma}}$ results in 
%changes of this ratio below 5\%.
%In the hadronic topologies, varying the cut on the jet-jet acoplanarity by 
%$\pm 5^\circ$ yields a change of less than 5\% in the Data/MC ratio.
%Changing the lower or upper limit in the jet-jet invariant mass window by 
%$\pm 10$~GeV results in a variation of the Data/MC ratio smaller than 2\%.  
%In the $jj\ell$ topology, the cut in the $\xi$ variable (see section~\ref{sec:single}) 
%gives a contribution not greater than 4\%,
%while changing the upper limit on the lepton energy by $\pm 5$~GeV results in a variation 
%of the Data/MC ratio smaller than 5\%. 
%Finally, the upper limit on the $\chi^2$ from the kinematic fits gives a 
%contribution smaller than 2\% in all hadronic topologies. 

An additional source of systematic error on the signal efficiency 
is due to the description of initial-state radiation (ISR) effects at
the event generator level.
Single excited-lepton events were generated with collinear ISR only.
As long as ISR photons are emitted below the DELPHI acceptance,
this results only in a small change in the event kinematics,
with low impact on the signal selection efficiency. 
However, if the ISR photon was detected, the event topology would be  
different from the topologies considered in the analysis, resulting in 
a smaller efficiency for the signal.
This effect was estimated using $e^+e^- \rightarrow e^+e^- $ and 
$e^+e^- \rightarrow \mu^+ \mu^-$ events simulated at  \cmes\ from 189~GeV  to 208~GeV.
For each \cme\ the ratio between the total number of generated events and the number of 
events with a photon with $E_\gamma > 5$~GeV  and 
$\theta_\gamma > 3^\circ$ was obtained.
By averaging over all \cmes\ an efficiency correction factor $k=0.90 \pm 0.02$ 
was obtained. 
The  error on $k$  was taken as an independent contribution to the systematic uncertainty of the signal efficiencies. 
%The fraction of events with an ISR photon of energy above 5~GeV
%(the minimum energy required for the candidates to isolated photons)  
%and with polar angle $\theta > 1.7^\circ$ was found to be of the order of 10\%. 

%The error on the luminosity measurement is another source of 
%systematic uncertainty affecting the total background prediction. 
%At DELPHI it is measured with a precision of the order of 1\% and 
%can thus be neglected.  

The systematic uncertainties discussed in this section were included in the
computation of the exclusion limits.

\subsection{Limits}
\label{sec:limits}

Limits at 95\% confidence level (CL) were computed using the modified frequentist 
likelihood ratio method described in reference~\cite{read}.
This method is well suited both for the combination of different search channels and for 
the inclusion of mass information. Searches in each topology at each \cme\ 
are treated as independent channels. 
Whenever the mass of the particle being searched for can be reconstructed, 
the PDF for a given signal mass hypothesis is assumed to be 
Gaussian with mean value equal to the tested signal mass and
standard deviation equal to the signal mass resolution. 
For channels where the excited-lepton  mass is not reconstructed, 
all selected events are  candidates for all signal mass hypotheses.
In the \mbox{$\ell^*\ell^*\rightarrow \nu\mathrm{W}\nu\mathrm{W}$} channel,
for which a discriminant analysis was performed, the PDFs of the likelihood ratio 
obtained at each signal mass hypothesis are used. 

\begin{figure}[p]
\begin{center}
\vspace{-1cm}
\mbox{\epsfig{file=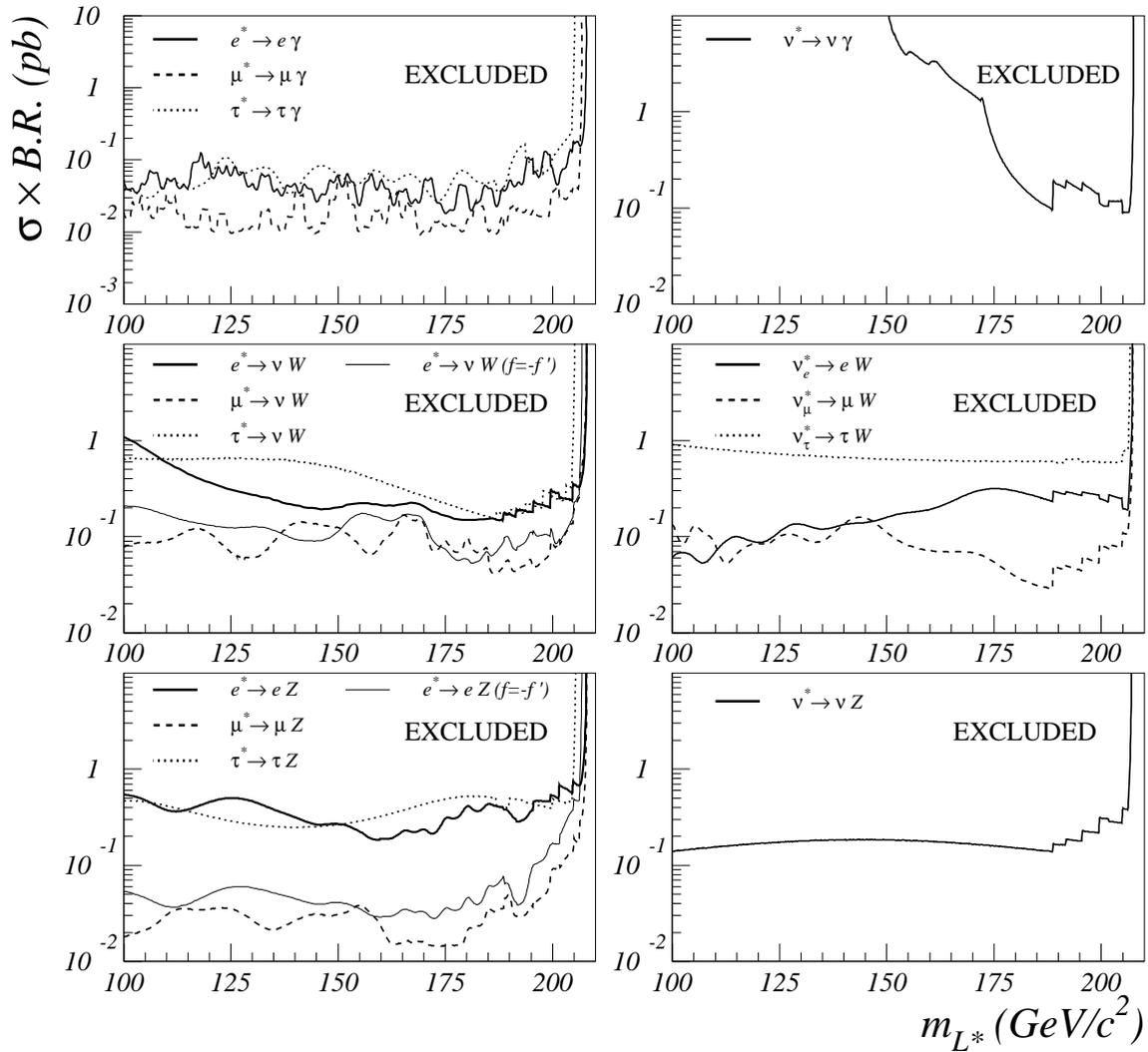,width=0.99\textwidth}}
\vspace{-2mm}
\caption[]
{Results on the single-production search of charged and neutral excited leptons.
The lines show the upper limits at 95\% CL
on $\sigma \times \mathrm{BR}$ as a function of the particle mass, for each
lepton flavour and decay mode.} 
\label{fig:lim_xsec}
\end{center}
\end{figure}

From the single-production search results, upper limits on the  
production cross-section multiplied by the  decay branching ratio 
($\sigma \times \mathrm{BR}$), as a function of the mass, were derived for 
each excited-lepton type and decay mode, as shown in figure~\ref{fig:lim_xsec}. 
As already discussed, the kinematics of single $e^*$ production is  
sensitive to the contribution from $t$-channel $\gamma$ exchange, with impact
on the selection efficiencies. 
The $e^*$  limits were thus computed using the selection efficiencies obtained 
with $f=f'$ and $f=-f'$. With $f=-f'$ the $e^*$ decay to a photon is forbidden.
For the other excited-lepton flavours the selection efficiencies do not depend 
on the $f$ and $f'$ assignments. The limits on $\sigma \times \mathrm{BR}$ can  
thus be interpreted in broader compositeness scenarios.

The cross-section for single excited-lepton production is a function not only of  
$m_{\elepton}$ but also of the coupling parameter $f/\Lambda$.   
The pair-production cross-section is a function of $m_{\elepton}$  only.
Upper limits on $f/\Lambda$  as a function of  $m_\elepton$ and
lower limits on  $m_\elepton$ were derived from the single- and pair-production 
searches respectively, by combining the results in the various decay modes.
The dependence of the decay BR's and production cross-sections on $m_{\elepton}$
as given in reference~\cite{boudjema} were assumed. The production cross-sections
were computed taking into account initial-state radiation effects.

\begin{figure}[p]
\begin{center}
\vspace{-1cm}
\mbox{\epsfig{file=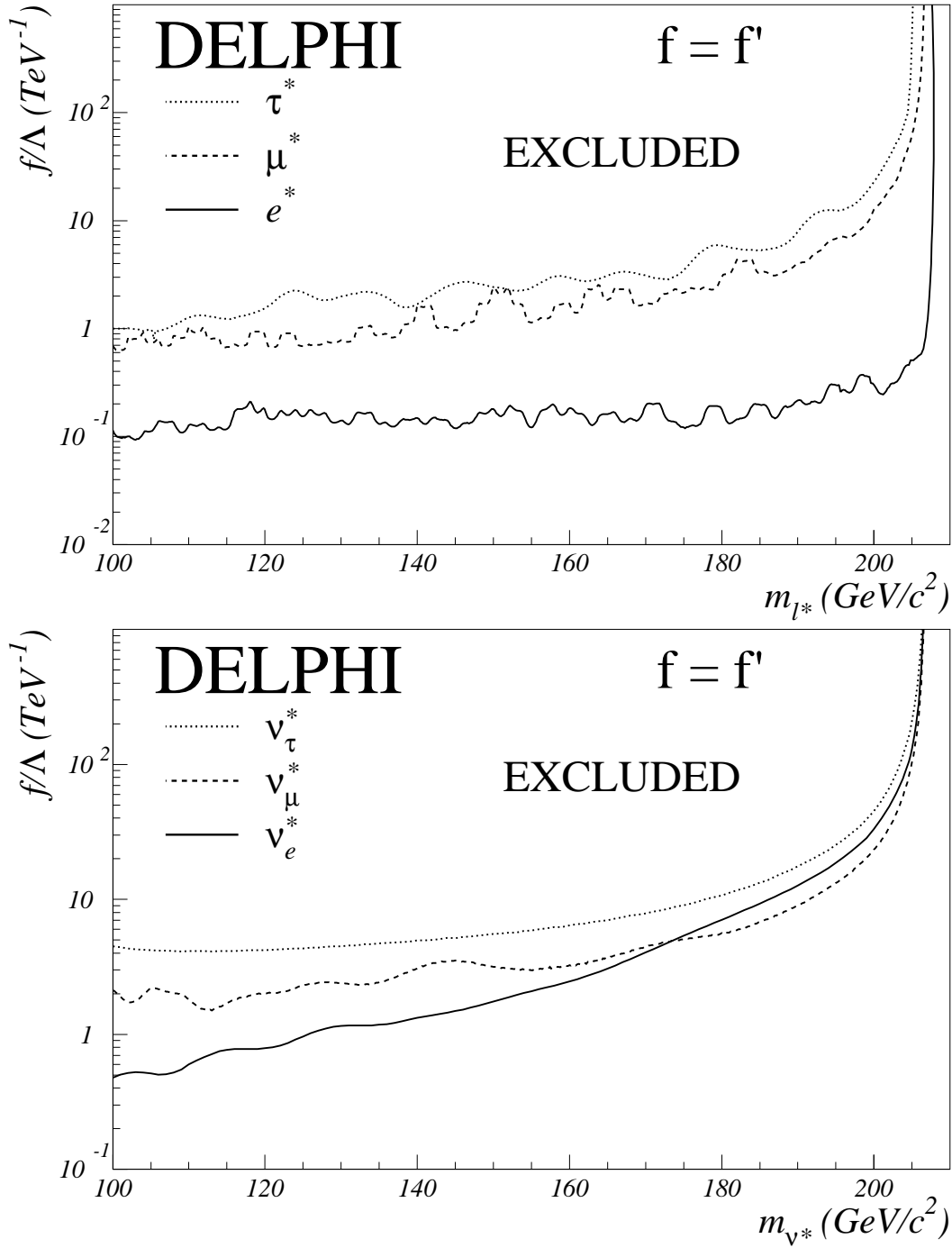,width=0.9\textwidth}}
\vspace{-2mm}
\caption[]
{ Results on single-production search of excited charged (upper plot) and neutral (lower plot)
leptons assuming $f=f'$. The lines show the 
upper limits at 95\% CL
on $f/\Lambda$ as a function of the excited-lepton mass.}
\label{fig:lim1}
\end{center}
\end{figure}

\begin{figure}[p]
\begin{center}
\vspace{-1cm}
\mbox{\epsfig{file=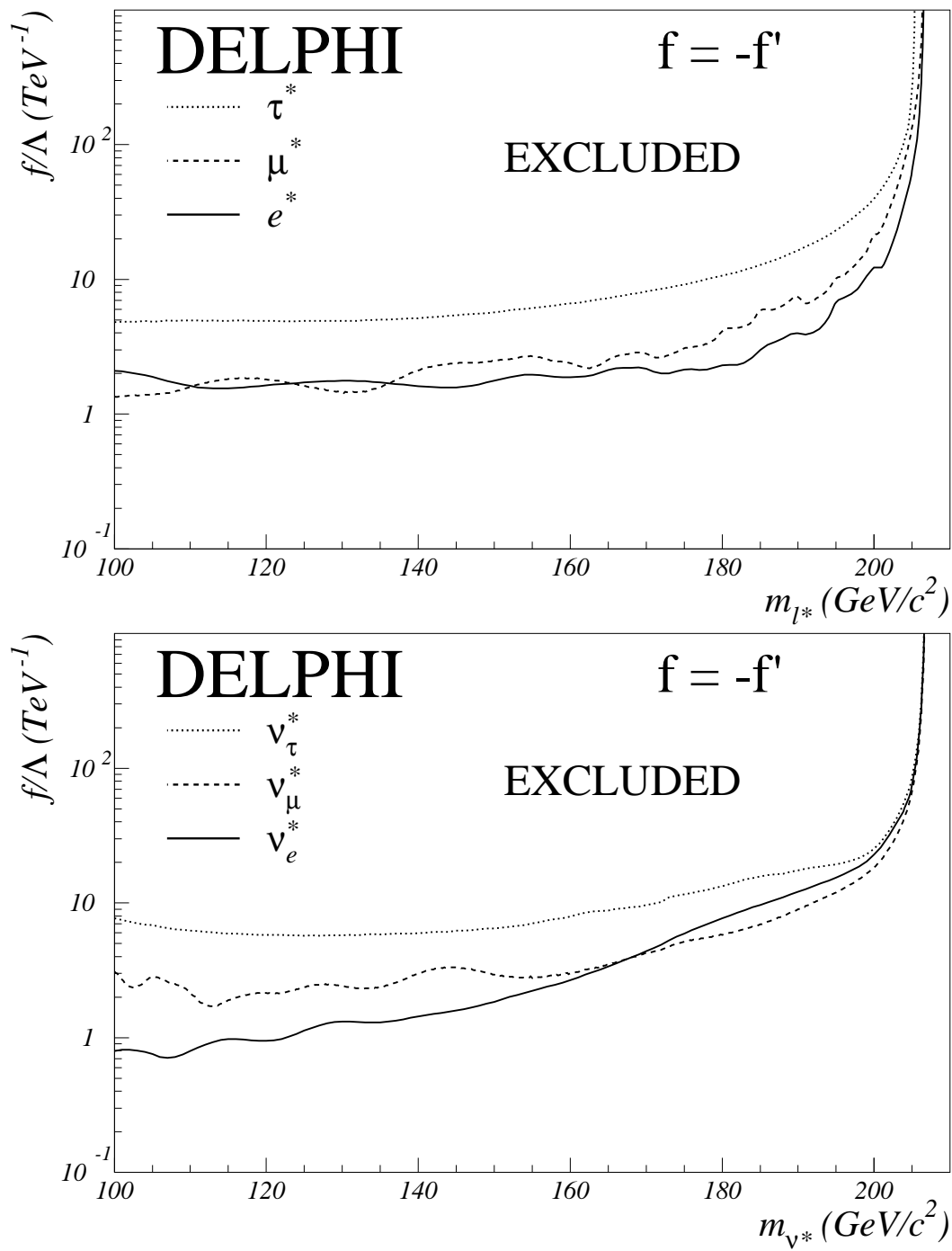,width=0.9\textwidth}}
\vspace{-2mm}
\caption[]{ As figure~\ref{fig:lim1}, but for  $f=-f'$.}
\label{fig:lim2}
\end{center}
\end{figure}

Figures~\ref{fig:lim1}  and \ref{fig:lim2} show the limits on 
$f/\Lambda$ as a function  of $m_\elepton $ for $f=f'$ and $f=-f'$, respectively.
The lower limits on the excited-lepton masses are given in table~\ref{tab:limits} 
for the two scenarios. It should be noted that the search for charged excited leptons 
yielded similar results in both scenarios. In the  $f=-f'$  case, although a poorer 
efficiency {\it vs} purity was obtained in the total number of selected events, the use
of the discriminant analysis was crucial to keep the signal sensitivity comparable
with the  $f=f'$ case.

\begin{figure}[p]
\begin{center}
\vspace{-1cm}
\mbox{\epsfig{file=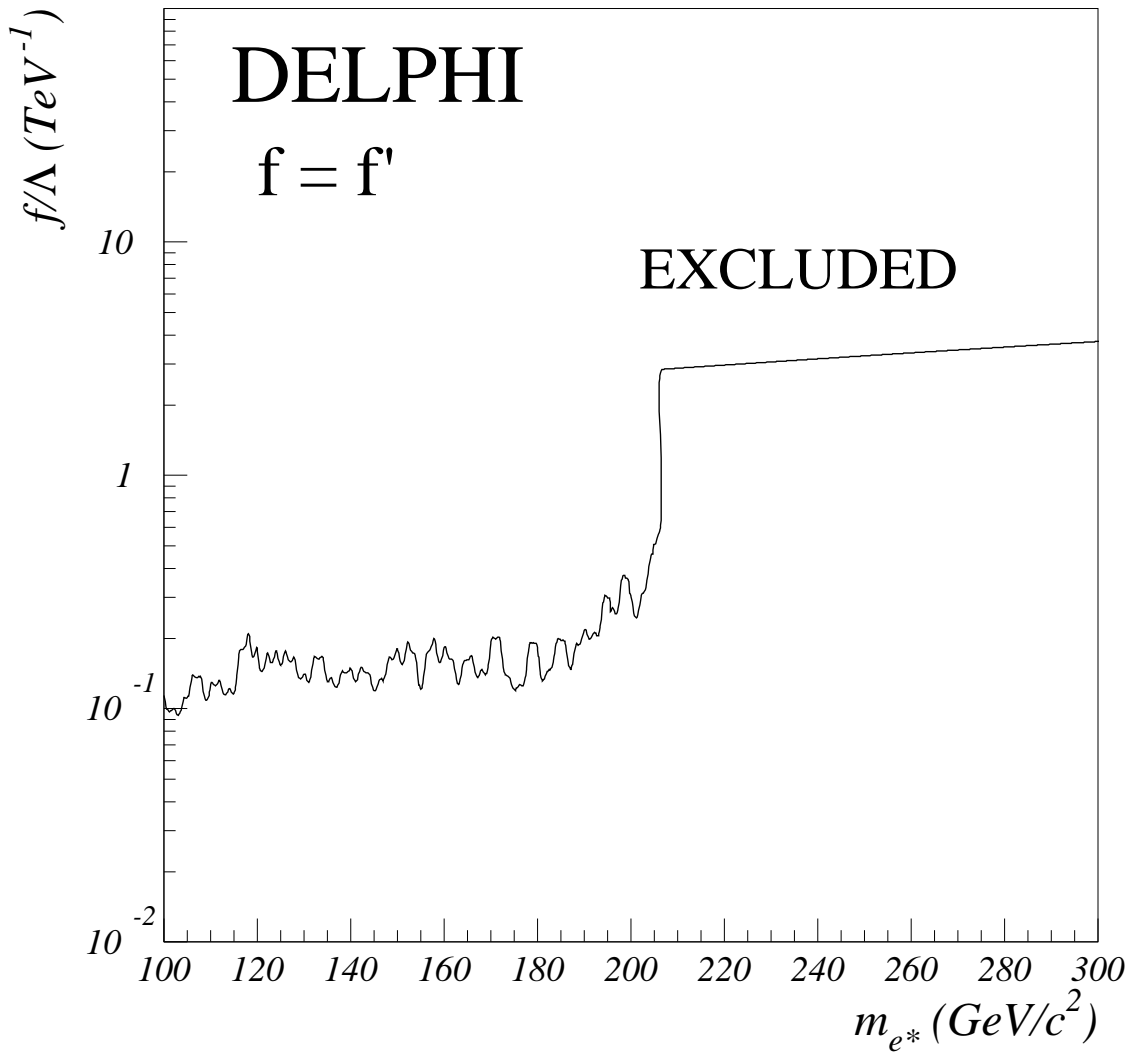,height=0.975\textwidth}}
\vspace{-0.75cm}
\caption[]
{Combined limit on excited electron production for $f=f'$ from direct and indirect 
searches. The line shows the upper limit at 95\% CL
on $f/\Lambda$. 
Up to the kinematic limit the result is dominated by the direct search for single 
production. For masses above the kinematic limit the result stems from the 
indirect search of excited electron contribution to the process 
\mbox{$e^+ e^- \rightarrow \gamma \gamma$} as described in reference~\cite{photons}.}
\label{fig:clim}
\end{center}
\end{figure}

Compositeness can also be probed at LEP through the process 
$e^+ e^- \rightarrow \gamma\gamma(\gamma)$. 
The additional contribution of the $t$-channel exchange of a virtual excited 
electron to the $e^+ e^- \rightarrow \gamma\gamma(\gamma)$ cross-section leads to a 
change in the angular distribution of the final-state photons with respect to the SM  
prediction. This effect depends on the excited electron mass $m_{e^*}$ and on the
$e e^* \gamma$ coupling. The results presented in reference~\cite{photons} were used 
to complement the direct searches for the excited electron in the mass 
region above the kinematic limit for $ee^*$ production. 
Figure~\ref{fig:clim} shows the upper limit on $f/\Lambda$ for the single production 
of excited electrons with $f=f'$, obtained by taking the best limit of the direct search
(figure~\ref{fig:lim1}(a)) and the indirect search results, 
thus extending the excluded region beyond the kinematic limit.

\section{Summary}
\label{sec:summary}

The data collected by DELPHI at $\sqrt{s}=189-209$~GeV, corresponding to 
an integrated luminosity of 598.7~pb$^{-1}$, were analysed to
search for excited leptons decaying promptly through
$\gamma$, \zboson\ or \wboson\ emission. 
No evidence for excited-lepton production was observed.
Limits on the model parameters were derived in two scenarios: $f=f'$ and $f=-f'$. 
From the single-production search  upper limits on $f/\Lambda$  as a function of 
$m_\elepton$ were set, as shown in Figures~\ref{fig:lim1}, \ref{fig:lim2} and 
\ref{fig:clim}. 
The search for pair production of excited leptons resulted in mass limits 
in the range \mbox{94 -- 103~GeV/$c^2$} depending on the excited-lepton type
and the assumed scenario for the coupling parameters, 
as quoted in table \ref{tab:limits}.
These limits are close to the kinematic limit for all charged (neutral) excited leptons
in the $f=f'$ ($f=-f'$) scenario.
Model independent upper bounds on $\sigma \times \mathrm{BR}$ 
were also derived for each excited-lepton flavour and decay channel, 
thus allowing for interpretations in broader compositeness scenarios.
%%%These results confirm and extend some of the limits set previously at LEP and HERA 
%%%\cite{delphi2,LEP,HERA}.  

%         Modified on 04-06-1999 by dimartino
%-------------------------------------------------------------------
\subsection*{Acknowledgements}
\vskip 3 mm
 We are greatly indebted to our technical 
collaborators, to the members of the CERN-SL Division for the excellent 
performance of the LEP collider, and to the funding agencies for their
support in building and operating the DELPHI detector.\\
We acknowledge in particular the support of \\
Austrian Federal Ministry of Education, Science and Culture,
GZ 616.364/2-III/2a/98, \\
FNRS--FWO, Flanders Institute to encourage scientific and technological 
research in the industry (IWT), Belgium,  \\
FINEP, CNPq, CAPES, FUJB and FAPERJ, Brazil, \\
Czech Ministry of Industry and Trade, GA CR 202/99/1362,\\
Commission of the European Communities (DG XII), \\
Direction des Sciences de la Mati$\grave{\mbox{\rm e}}$re, CEA, France, \\
Bundesministerium f$\ddot{\mbox{\rm u}}$r Bildung, Wissenschaft, Forschung 
und Technologie, Germany,\\
General Secretariat for Research and Technology, Greece, \\
National Science Foundation (NWO) and Foundation for Research on Matter (FOM),
The Netherlands, \\
Norwegian Research Council,  \\
State Committee for Scientific Research, Poland, SPUB-M/CERN/PO3/DZ296/2000,
SPUB-M/CERN/PO3/DZ297/2000, 2P03B 104 19 and 2P03B 69 23(2002-2004),\\
FCT - Funda\c{c}\~ao para a Ci\^encia e Tecnologia, Portugal, \\
Vedecka grantova agentura MS SR, Slovakia, Nr. 95/5195/134, \\
Ministry of Science and Technology of the Republic of Slovenia, \\
CICYT, Spain, AEN99-0950 and AEN99-0761,  \\
The Swedish Research Council,      \\
Particle Physics and Astronomy Research Council, UK, \\
Department of Energy, USA, DE-FG02-01ER41155, \\
EEC RTN contract HPRN-CT-00292-2002. \\

%=========================================================================%

\newpage
%%%%%%%%%%%%%%%%%%%%%%%%%%%%%%%%%%%%%%%%%%%%%%%%%%%%%%% tables
{\small
\begin{table}[hbt]
\begin{center}
{
\vskip -0.4 cm
\begin{tabular}{|l|l|rl|rl|rl|}
\hline
$\sqrt{s}$ &  & \multicolumn{6}{c|}{Excited-lepton flavour }  
\\
\cline{3-8} 
(GeV)  & Channel  & \multicolumn{2}{c|}{$e$} &   \multicolumn{2}{c|}{$\mu$}  & 
 \multicolumn{2}{c|}{$\tau$} \\
\hline
\hline
    & $\ell^*\rightarrow\ell\gamma$   &366&(408.2$\pm$6.7)&39&(44.0$\pm$1.5)&62&(54.3$\pm$2.7)
\\ \cline{2-8}
    & $\ell^*\rightarrow\nu\wboson$   & 202 & (217.6$\pm$4.8)   & 195  & (195.5$\pm$4.3)   & 432 & (447.8$\pm$7.1) 
\\ \cline{2-8}
189 & $\ell^*\rightarrow\ell\zboson$  &  51 & (42.0$\pm$3.4)    &   8  & (10.00$\pm$0.66)  &  55 & (40.9$\pm$2.7)  
\\ \cline{2-8}
    & $\nu^*\rightarrow\nu \gamma$    &      \multicolumn{6}{c|}{3 (0.70$\pm$0.17)}
\\ \cline{2-8}
    & $\nu^*\rightarrow\ell\wboson$   & 175 & (180.1$\pm$4.9)   & 133  & (131.3$\pm$3.2)   & 203  & (196.5$\pm$4.9)
\\ \cline{2-8}
    & $\nu^*\rightarrow\nu\zboson$    &      \multicolumn{6}{c|}{75 (81.1$\pm$3.1)}
\\ \cline{2-8}

\hline
\hline
    & $\ell^*\rightarrow\ell\gamma$   &  61 & ( 66.4$\pm$2.0)    &   8 & (  7.29$\pm$0.27) & 10 & (  8.05$\pm$0.76)    
\\ \cline{2-8}
    & $\ell^*\rightarrow\nu\wboson$   &  28 & ( 38.04$\pm$0.79)    &  25 & ( 33.71$\pm$0.69)    &  56 & ( 75.2$\pm$1.1)    
\\ \cline{2-8}
192 & $\ell^*\rightarrow\ell\zboson$  &   5 & (  5.24$\pm$0.44)    &   1 & (  2.01$\pm$0.16)    &   3 & (  6.56$\pm$0.43)    
\\ \cline{2-8}
    & $\nu^*\rightarrow\nu \gamma$    &      \multicolumn{6}{c|}{0 (0.10$\pm$0.03)}
\\ \cline{2-8}
    & $\nu^*\rightarrow\ell\wboson$   &  24 & ( 28.53$\pm$0.75)    &  18 & ( 22.01$\pm$0.53)    &  26 & ( 31.33$\pm$0.79)    
\\ \cline{2-8}
    & $\nu^*\rightarrow\nu\zboson$    &   \multicolumn{6}{c|}{9 ( 14.87$\pm$0.49) }
\\ \cline{2-8}

\hline
\hline
    & $\ell^*\rightarrow\ell\gamma$   & 201 & (188.9$\pm$3.3)    &  18 & (20.27$\pm$0.73)    &  24 & (23.4$\pm$1.3)    
\\ \cline{2-8}
    & $\ell^*\rightarrow\nu\wboson$   & 108 & (117.3$\pm$2.4)    & 103 & (101.6$\pm$2.1)    & 232 & (230.0$\pm$3.5)    
\\ \cline{2-8}
196 & $\ell^*\rightarrow\ell\zboson$  &  26 & (20.4$\pm$1.5)    &   3 & (5.81$\pm$0.36)    &  23 & (21.0$\pm$1.2)    
\\ \cline{2-8}
    & $\nu^*\rightarrow\nu \gamma$    &      \multicolumn{6}{c|}{2 (0.70$\pm$0.08)}
\\ \cline{2-8}
    & $\nu^*\rightarrow\ell\wboson$   &  90 & (91.2$\pm$2.3)    &  72 & (64.6$\pm$1.6)    & 106 & (96.7$\pm$2.4)    
\\ \cline{2-8}
    & $\nu^*\rightarrow\nu\zboson$    &  \multicolumn{6}{c|}{ 38 (45.6$\pm$1.5)}
\\ \cline{2-8}

\hline
\hline
    & $\ell^*\rightarrow\ell\gamma$   & 190 & (198.8$\pm$3.5)    &  25 & (20.74$\pm$0.74)    &  26 & (27.3$\pm$1.3)    
\\ \cline{2-8}
    & $\ell^*\rightarrow\nu\wboson$   & 128 & (131.1$\pm$2.6)    & 102 & (114.6$\pm$2.4)    & 239 & (248.0$\pm$3.7)    
\\ \cline{2-8}
200 & $\ell^*\rightarrow\ell\zboson$  &  29 & (19.3$\pm$1.4)    &   9 & ( 7.05$\pm$0.48)    &  28 & (22.4$\pm$1.3)    
\\ \cline{2-8}
    & $\nu^*\rightarrow\nu \gamma$    &      \multicolumn{6}{c|}{6 (1.7$\pm$0.1)}
\\ \cline{2-8}
    & $\nu^*\rightarrow\ell\wboson$   & 116 & (96.7$\pm$2.4)    &  70 & (70.5$\pm$1.7)    & 108 & (102.5$\pm$2.5)    
\\ \cline{2-8}
    & $\nu^*\rightarrow\nu\zboson$    &  \multicolumn{6}{c|}{42 (52.2$\pm$1.7) }
\\ \cline{2-8}

\hline 
\hline 
    & $\ell^*\rightarrow\ell\gamma$   &  87 & (94.9$\pm$2.1)    &   8 & (9.44$\pm$0.38)    &  16 & (12.05$\pm$0.73)    
\\ \cline{2-8}
    & $\ell^*\rightarrow\nu\wboson$   &  79 & (61.9$\pm$1.2)    &  54 & (54.1$\pm$1.1)    & 138 & (119.6$\pm$1.8)    
\\ \cline{2-8}
202 & $\ell^*\rightarrow\ell\zboson$  &   5 & (10.53$\pm$0.77)    &   4 & (3.45$\pm$0.25)    &   9 & (11.42$\pm$0.68)    
\\ \cline{2-8}
    & $\nu^*\rightarrow\nu \gamma$    &      \multicolumn{6}{c|}{1 (1.2$\pm$0.1)}
\\ \cline{2-8}
    & $\nu^*\rightarrow\ell\wboson$   &  54 & (46.8$\pm$1.2)    &  28 & (33.57$\pm$0.84)    &  55 & (51.4$\pm$1.3)    
\\ \cline{2-8}
    & $\nu^*\rightarrow\nu\zboson$    &  \multicolumn{6}{c|}{31  (24.91$\pm$0.80)     }
\\ \cline{2-8}

\hline
\hline
    & $\ell^*\rightarrow\ell\gamma$   & 179 & (182.7$\pm$3.8)    &  20 & (18.28$\pm$0.34)    &  27 & (24.3$\pm$1.2)    
\\ \cline{2-8}
    & $\ell^*\rightarrow\nu\wboson$   & 119 & (123.8$\pm$2.4)    &  94 & (102.4$\pm$2.1)    & 225 & (231.5$\pm$3.4)    
\\ \cline{2-8}
205 & $\ell^*\rightarrow\ell\zboson$  &  22 & (18.0$\pm$1.3)    &   7 & (6.82$\pm$0.47)    &  24 & (21.5$\pm$1.3)    
\\ \cline{2-8}
    & $\nu^*\rightarrow\nu \gamma$    &      \multicolumn{6}{c|}{5 (1.3$\pm$0.2)}
\\ \cline{2-8}
    & $\nu^*\rightarrow\ell\wboson$   &  98 & (90.9$\pm$2.3)    &  60 & (61.8$\pm$1.6)    &  88 & (100.1$\pm$2.4)    
\\ \cline{2-8}
    & $\nu^*\rightarrow\nu\zboson$    &  \multicolumn{6}{c|}{ 46 (50.5$\pm$1.6) }
\\ \cline{2-8}

\hline
\hline
    & $\ell^*\rightarrow\ell\gamma$   & 120 & (129.8$\pm$2.7)    &  11 & (13.15$\pm$0.25)    &  16 & (19.2$\pm$1.0)    
\\ \cline{2-8}
    & $\ell^*\rightarrow\nu\wboson$   &  79 & (93.9$\pm$1.8)    &  70 & (79.8$\pm$1.7)    & 138 & (173.1$\pm$2.5)    
\\ \cline{2-8}
206$^{*}$ & $\ell^*\rightarrow\ell\zboson$  &  15 & (14.3$\pm$1.0)    &   3 & (4.51$\pm$0.35)    &  16 & (14.75$\pm$0.88)    
\\ \cline{2-8}
    & $\nu^*\rightarrow\ell\wboson$   &  58 & (67.4$\pm$1.7)    &  41 & (46.4$\pm$1.2)    &  53 & (72.3$\pm$1.7)  \\ \cline{2-8}
    & $\nu^*\rightarrow\nu\zboson$    &  \multicolumn{6}{c|}{  37 (38.9$\pm$1.2)  }
\\ \cline{2-8}

\hline
\hline
$> 206$    & $\nu^*\rightarrow\nu \gamma$    &      \multicolumn{6}{c|}{1 (2.6$\pm$0.2)}
\\ \cline{2-8}

\hline
\hline
    & $\ell^*\rightarrow\ell\gamma$   & 172 & (180.9$\pm$3.3)    &  15 & (18.28$\pm$0.34)    &  21 & (25.1$\pm$1.2)    
\\ \cline{2-8}
    & $\ell^*\rightarrow\nu\wboson$   & 110 & (130.6$\pm$2.5)    &  94 & (110.7$\pm$2.3)    & 205 & (239.3$\pm$3.5)    
\\ \cline{2-8}
207 & $\ell^*\rightarrow\ell\zboson$  &  30 & (22.8$\pm$1.6)    &   6 & (6.9$\pm$0.5)    &  21 & (23.4$\pm$1.3)    
\\ \cline{2-8}
$+$    & $\nu^*\rightarrow\ell\wboson$   &  86 & (96.4$\pm$2.4)    &  52 & (64.9$\pm$1.7)    &  87 & (99.9$\pm$2.4)    
\\ \cline{2-8}
208    & $\nu^*\rightarrow\nu\zboson$    &  \multicolumn{6}{c|}{  53 (55.3$\pm$1.7)  }
\\ \cline{2-8}

\hline 

\end{tabular}

}
\end{center}

\vskip -0.4 cm
\caption[]{Number of candidates for the different excited-lepton decay channels 
in the single-production search.
The numbers in parentheses correspond to the SM background expectations with the
statistical errors.}
\label{tab:single}
\end{table}
}
%%%%%%%%%%%%%%%%%%%%%%%%%%%%%%%%%%%%%%%%%%%%%%%%%%%%%%%%%%%%%%%%%%%%%%%%%%%%%%%%%%
\normalsize
%%%%%%%%%%%%%%%%%%%%%%%%%%%%%%%%%%%%%%%%%%%%%%%%%%%%%%%%%%%%%%%%%%%%%%%%%%%%%%%%%%
\begin{table}[hbt]
\begin{center}
%\vskip 0.45 cm
{
\begin{tabular}{|c|l|c|c|c|}
\hline
$\sqrt{s}$ &  & \multicolumn{3}{c|}{Excited-lepton flavour }  \\
\cline{3-5} 
(GeV)  & Channel  & $e$ & $\mu$  & $\tau$                   \\
\hline
\hline

 & $\ell^* \rightarrow \ell\gamma$ & 0 (1.25$\pm$0.33) & 0 (0.38$\pm$0.05 ) & 1 (2.57$\pm$0.41) 

\\ 
\cline{2-5}
    & $\ell^* \rightarrow \nu W$ (hadr.)    & \multicolumn{3}{c|}{403 (419.9$\pm$5.1)} 
\\ 
%\cline{2-5}
    & $\ell^* \rightarrow \nu W$ (semilep.) & \multicolumn{3}{c|}{261 (231.1$\pm$3.1)} 
\\ 
\cline{2-5}
205 & $\nu^* \rightarrow \nu\gamma$   & \multicolumn{3}{c|}{4 (3.26$\pm$0.33)}  
\\ 
\cline{2-5}
 & $\nu^* \rightarrow \ell W$  & 1 (3.21$\pm$0.33) & 1 (0.99$\pm$0.16) & 18 (17.43$\pm$0.87)      
\\ 
\hline
\hline
%%%%%%%%%%%%%%%%%%%%%%%%%%%%%%%%%%%%%%%%%%%%%%%%%%%%%%%%%%%%%%%%%%%%%%%%%%%%%%%%
 & $\ell^* \rightarrow \ell\gamma$ &1 (0.52$\pm$0.20)& 0 (0.29$\pm$0.04) & 3 (1.48$\pm$0.27) 
\\ 
\cline{2-5}
  & $\ell^* \rightarrow \nu W$ (hadr.)      & \multicolumn{3}{c|}{280 (311.0$\pm$3.8)}
\\ 
%\cline{2-5}
    & $\ell^* \rightarrow \nu W$ (semilep.) & \multicolumn{3}{c|}{154 (176.6$\pm$2.4)}
\\ 
\cline{2-5}
206$^{*}$ & $\nu^* \rightarrow \nu\gamma$   & \multicolumn{3}{c|}{6 (2.21$\pm$0.26)}  
\\ 
\cline{2-5}
 & $\nu^* \rightarrow \ell W$  & 6 (2.56$\pm$0.32) & 0 (0.93$\pm$0.14) & 15 (12.04$\pm$0.61)   
\\ 
\hline
\hline
%%%%%%%%%%%%%%%%%%%%%%%%%%%%%%%%%%%%%%%%%%%%%%%%%%%%%%%%%%%%%%%%%%%%%%%%%%%%%%%%
 & $\ell^* \rightarrow \ell\gamma$ &1 (1.49$\pm$0.34)& 2 (0.41$\pm$0.05) & 3 (2.59$\pm$0.39) 
\\ 
\cline{2-5}
  & $\ell^* \rightarrow \nu W$ (hadr.)      & \multicolumn{3}{c|}{408 (416.9$\pm$5.0)}
\\ 
%\cline{2-5}
    & $\ell^* \rightarrow \nu W$ (semilep.) & \multicolumn{3}{c|}{228 (239.0$\pm$3.2)}
\\ 
\cline{2-5}
207 & $\nu^* \rightarrow \nu\gamma$   & \multicolumn{3}{c|}{3 (3.50$\pm$0.36)}  
\\ 
\cline{2-5}
 & $\nu^* \rightarrow \ell W$  & 3 (3.10$\pm$0.32) & 1 (1.74$\pm$0.22) & 20 (17.22$\pm$0.89)   
\\ 
\hline
\hline
%%%%%%%%%%%%%%%%%%%%%%%%%%%%%%%%%%%%%%%%%%%%%%%%%%%%%%%%%%%%%%%%%%%%%%%%%%%%%%%%
 & $\ell^* \rightarrow \ell\gamma$ &0 (0.16$\pm$0.06)& 0 (0.04$\pm$0.01) & 0 (0.25$\pm$0.07) 
\\  
\cline{2-5}
 & $\ell^* \rightarrow \nu W$ (hadr.)       & \multicolumn{3}{c|}{34 (34.95$\pm$0.85)}
\\ 
%\cline{2-5}
    & $\ell^* \rightarrow \nu W$ (semilep.) & \multicolumn{3}{c|}{10 (19.94$\pm$0.54)}
\\ 
\cline{2-5}
208  & $\nu^* \rightarrow \nu\gamma$   & \multicolumn{3}{c|}{-}  
\\
\cline{2-5}
  & $\nu^* \rightarrow \ell W$  &  0 (0.36$\pm$0.08) & 0 (0.10$\pm$0.02) & 2 (1.68$\pm$0.17)      
   
\\ 
\hline
\end{tabular}
}
\end{center}
\vskip -0.4 cm
\caption[]{Number of excited-lepton candidates for the different decay channels
and centre-of-mass energies in the pair-production search. 
The numbers in parentheses correspond to the SM background expectations with the
statistical errors.}
\label{tab:pair}
\end{table}

%%%%%%%%%%%%%%%%%%%%%%%%%%%%%%%%%%%%%%%%%%%%%%%%%%%%%%%%%%%%%%%%%%%%%%%%%%%%%%%%%%
\begin{table}[hbt]
\begin{center}
{
\vskip -0.4 cm
\begin{tabular}{|l|rl|rl|rl|}
\hline
 & \multicolumn{6}{|c|}{Excited lepton flavour }  
\\
\cline{2-7} 
 Channel  & \multicolumn{2}{|c|}{$e$} &   \multicolumn{2}{|c|}{$\mu$}  & 
 \multicolumn{2}{|c|}{$\tau$} \\
\hline
   \multicolumn{7}{|c|}{\hspace{3.5cm} Single production} \\
\hline
   $\ell^*\rightarrow\ell\gamma$   & 1376 & (1451 $\pm$ 10)   & 144 & (151.5 $\pm$ 2.0) &  202 & (193.7 $\pm$ 4.0)
\\ \cline{1-7}
   $\ell^*\rightarrow\nu\wboson$   &  853 & (914.2 $\pm$ 7.3) & 737 & (792.4 $\pm$ 6.5) & 1665 & (1765$\pm$ 11) 
\\ \cline{1-7}
   $\ell^*\rightarrow\ell\zboson$  &  183 & (152.6 $\pm$ 4.7) &  41 & (46.5 $\pm$ 1.2) &  179 & (161.9 $\pm$ 3.9)  
\\ \cline{1-7}
   $\nu^*\rightarrow\nu \gamma$    &      \multicolumn{6}{c|}{ 18 (8.3 $\pm$ 0.4)}
\\ \cline{1-7}
   $\nu^*\rightarrow\ell\wboson$   &  701 & (698.0 $\pm$ 7.1)   & 474 & (495.1 $\pm$ 4.9) & 726 & (750.7 $\pm$ 7.3)
\\ \cline{1-7}
   $\nu^*\rightarrow\nu\zboson$    &      \multicolumn{6}{c|}{331 (363.4 $\pm$ 4.7)}
\\ \cline{1-7}

\hline
  \multicolumn{7}{|c|}{\hspace{3.5cm} Pair production} \\
\hline
 $\ell^* \rightarrow \ell\gamma$ & 2 & (3.42 $\pm$ 0.52) & 2 & (1.12 $\pm$ 0.08) & 7 & (6.89 $\pm$0.63) \\ 
\cline{1-7}
 $\ell^* \rightarrow \nu W$ (hadr.)    & \multicolumn{6}{c|}{1125 (1182.7 $\pm$ 8.1)} 
\\ 
 $\ell^* \rightarrow \nu W$ (semilep.) & \multicolumn{6}{c|}{653 (666.6 $\pm$ 5.1)} 
\\ 
\cline{1-7}
 $\nu^* \rightarrow \nu\gamma$   & \multicolumn{6}{c|}{ 13 (8.97 $\pm$ 0.55)}  
\\ 
\cline{1-7}
 $\nu^* \rightarrow \ell W$  &  10 & (9.23 $\pm$ 0.57) & 2 & (3.76 $\pm$ 0.31) & 55 & (48.4 $\pm$ 1.4)      
\\ 

\hline

\end{tabular}

}
\end{center}

\vskip -0.4 cm
\caption[]{Total number of candidates for the different excited lepton decay channels 
in the single (top) and pair (bottom) production searches.
The numbers in parentheses correspond to the SM background expectations with the
statistical errors.}
\label{tab:all}
\end{table}

%%%%%%%%%%%%%%%%%%%%%%%%%%%%%%%%%%%%%%%%%%%%%%%%%%%%%%%%%%%%%%%%%%%%%%%%%%%%%%%%%%

%%%%%%%%%%%%%%%%%%%%%%%%%%%%%%%%%%%%%%%%%%%%%%%%%%%%%%%%%%%%%%%%%%%%%%%%%%%%%%%%%%
\vspace{ -2.4 cm}
%%%%%%%%%%%%%%%%%%%%%%%%%%%%%%%%%%%%%%%%%%%%%%%%%%%%%

%%%%%%%%%%%%%%%%%%%%%%%%%%%%%%%%%%%%%%%%%%%%%%%%%%%%%%%%%%%%%%%%%%%%%%%%%%%%%%%%%%
\begin{table}[hbt]

\begin{center}
\begin{tabular}{|l|c|c|c|}
\hline
  & \multicolumn{3}{c|}{Excited lepton flavour}  
\\
\cline{2-4} 
channel  & $e$ &  $\mu$  & $\tau$ 
\\
\hline
\hline
\multicolumn{4}{|c|}{\hspace{0.7cm} Single production} \\
\hline    
 $\ell^* \rightarrow \ell\gamma$ & 0.39, 0.43, 0.40  & 0.58, 0.58, 0.61  & 0.29, 0.30, 0.23  \\ 
\hline
 $\ell^* \rightarrow \nu W$      & 0.20, 0.24, 0.33  & 0.32, 0.32, 0.33  & 0.24, 0.25, 0.43    \\ 
\hline 
 $\ell^* \rightarrow \ell Z$     & 0.05, 0.04, 0.20  & 0.39, 0.39, 0.42  & 0.16, 0.16, 0.20   \\ 
\hline 
 $\nu^* \rightarrow \nu \gamma$  & \multicolumn{3}{c|}{ 0.0, 0.0, 0.37 }                          \\ 
\hline 
 $\nu^* \rightarrow \ell W$      & 0.27, 0.31, 0.34 &  0.27, 0.34, 0.34  &  0.16, 0.19, 0.18 \\ 
\hline 
 $\nu^* \rightarrow \nu Z $      & \multicolumn{3}{c|}{0.30, 0.31, 0.41}                     \\ 
\hline
\multicolumn{4}{|c|}{\hspace{0.7cm} Pair production} \\
\hline
$\ell^* \rightarrow \ell\gamma$ & 0.38  & 0.51 & 0.17  
\\ 
\hline
$\ell^* \rightarrow \nu W$ (hadr.)    &  \multicolumn{3}{c|}{0.22} 
\\  
$\ell^* \rightarrow \nu W$ (semilep.) &  \multicolumn{3}{c|}{0.16} 
\\ 
\hline
$\nu^* \rightarrow \nu\gamma$   &  \multicolumn{3}{c|}{0.53} 
\\ 
\hline
$\nu^* \rightarrow \ell W$ & 0.28 & 0.42 & 0.13   
\\ 
\hline
\end{tabular}
\end{center}

\vskip -0.4 cm

\caption[]{Selection efficiencies  for the different excited lepton flavours 
and decay channels, in  the single (top) and pair (bottom) production modes. 
Efficiencies are quoted for excited lepton masses of $m_{\elepton}$=125, 150 and 200~GeV/$c^2$ 
in the single production and  $m_{\elepton}$=100~GeV/$c^2$ in the pair production,
at $\sqrt{s}=$206~GeV. The relative statistical errors range between 3\% and 8\%, 
depending on the channel.}
\label{tab:effi}
\end{table}

%%%%%%%%%%%%%%%%%%%%%%%%%%%%%%%%%%%%%%%%%%%%%%%%%%%%%%%%%%%%%%%%%%%%%%%%%%%%%%%%%%

\begin{table}[htb]
\begin{center}
\begin{tabular}{|c||c|c|c|} \hline
         &   $e^*$   &  $\mu^*$   &   $\tau^*$    \\ 
\hline
\hline
$f=f'$     & 103.1  & 103.2 & 102.7 \\ \hline
$f=-f'$    & 101.0  & 101.0 & 101.0 \\ \hline 
\end{tabular}
\begin{tabular}{|c||c|c|c|} \hline
    & $\nu^*_{e}$    &   $\nu^*_{\mu}$  &     $\nu^*_{\tau}$ \\
\hline
\hline
$f=f'$     & 
101.9 & 103.2 & 94.2 \\ \hline
$f=-f'$ & 
101.9 & 102.2 & 101.9 \\ \hline
\end{tabular}
\end{center}
\vskip -0.4 cm
\caption[]{Lower limits (in GeV/$c^2$) at 95 \% CL on the
excited-lepton masses obtained from the pair-production searches.}
\label{tab:limits}
\end{table}

%%%%%%%%%%%%%%%%%%%%%%%%%%%%%%%%%%%%%%%%%%%%%%%%%%%%%% figures
\newpage
%%%%%%%%%%%%%%%%%%%%%%%%%%%%%%%%%%%%%%%%%%%%%%%%%%%%%%%%%%%%%%%%%%%%%%%%%%%%%%%%%%

%%%%%%%%%%%%%%%%%%%%%%%%%%%%%%%%%%%%%%%%%%%%%%%%%%%%%%% 
\newpage

%%%%%%%%%%%%%%%%%%%%%%%%%%%%%%%%%%%%%%%%%%%%%%%%%%%%%%%%%%%%%%%%%%%%%%%%%
\newpage

%%%%%%%%%%%%%%%%%%%%%%%%%%%%%%%%%%%%%%%%%%%%%%%%%%%%%%%%%%%%%%%%%%%%%%%%%

%%%%%%%%%%%%%%%%%%%%%%%%%%%%%%%%%%%%%%%%%%%%%%%%%%%%%%%%%%%%%%%%%%%%%%%%%
\newpage

%%%%%%%%%%%%%%%%%%%%%%%%%%%%%%%%%%%%%%%%%%%%%%%%%%%%%%%%%%%%%%%%%%%%%%%%%

%\newpage

%%%%%%%%%%%%%%%%%%%%%%%%%%%%%%%%%%%%%%%%%%%%%%%%%%%%%%%%%%%%%%%%%%%%%%%%%

\newpage

%%%%%%%%%%%%%%%%%%%%%%%%%%%%%%%%%%%%%%%%%%%%%%%%%%%%%%%%%%%%%%%%%%%%%%%%%
\newpage

%%%%%%%%%%%%%%%%%%%%%%%%%%%%%%%%%%%%%%%%%%%%%%%%%%%%%%%%%%%%%%%%%%%%%%%%%
\newpage

\newpage

\newpage

\end{document}